\documentstyle[aps,eqsecnum]{revtex}
\begin{document}
\draft
\title{Gravitational radiation from a particle
       in circular orbit around a black hole. \\
       V. Black-hole absorption and tail corrections}
\author{Eric Poisson}
\address{McDonnell Center for the Space Sciences,
         Department of Physics, Washington University,
         St.~Louis, Missouri 63130}
\author{Misao Sasaki\thanks{Present address: Department of
        Earth and Space Science, Osaka University, Toyonaka,
        Osaka 560, Japan.}}
\address{Department of Physics, Faculty of Science,
         Kyoto University, Kyoto 606, Japan}
\date{draft \today}
\maketitle
\begin{abstract}
A particle of mass $\mu$ moves on a circular orbit of
a nonrotating black hole of mass $M$. Under the
restrictions $\mu/M \ll 1$ and $v \ll 1$, where $v$
is the orbital velocity (in units in which $c=1$),
we consider the gravitational
waves emitted by such a binary system. The framework
is that of black-hole perturbation theory.
We calculate $\dot{E}$, the rate at which the gravitational
waves remove energy from the system. The total energy loss is
given by $\dot{E} = \dot{E}^\infty + \dot{E}^H$, where
$\dot{E}^\infty$ denotes that part of the gravitational-wave
energy which is carried off to infinity, while $\dot{E}^H$
denotes the part which is absorbed by the black hole. We
show that the black-hole absorption is a small effect:
$\dot{E}^H/\dot{E} \simeq v^8$. This is
explained by the presence of a potential barrier in the
vicinity of the black hole: Most of the waves propagating
initially toward the black hole are reflected off the
barrier; the black hole is therefore unable to absorb
much. The black-hole absorption (and indeed any other effect
resulting from imposing ingoing-wave boundary conditions
at the event horizon) are sufficiently small to be
irrelevant to the construction of matched filters for
gravitational-wave measurements. To derive this result
we extend the techniques previously developed by Poisson
and Sasaki for integrating the Regge-Wheeler equation. The
extension consists of an explicit consideration of the
horizon boundary conditions, which were largely ignored
in the previous work. Finally, we compare
the wave generation formalism which derives from perturbation
theory to the post-Newtonian formalism of Blanchet and Damour.
Among other things we consider the corrections to the asymptotic
gravitational-wave field which are due to wave-propagation
(tail) effects. The results obtained using perturbation theory
are identical to that of post-Newtonian theory.
\end{abstract}
\twocolumn
\narrowtext

\section{Introduction and summary}

\subsection{Gravitational waves from coalescing compact
            binaries}

Coalescing compact binary systems,
composed of neutron stars and/or
black holes, have been identified as the most promising
source of gravitational waves for kilometer-size
interferometric detectors \cite{Thorne1987,Schutz}.

The construction of three
such detectors should be completed by the turn of the
century. The American LIGO (Laser Interferometer
Gravitational-wave Observatory) project \cite{LIGO}
involves two
4 km detectors, one situated in Hanford, Washington,
the other in Livingston, Louisiana; construction has
begun at the Hanford cite. The French-Italian Virgo
project \cite{Virgo}
involves a single 3 km detector, to be built
near Pisa, Italy.

According to recent estimates
\cite{Narayanetal,Phinney}, these interferometers
should be able to detect approximately three binary
coalescences per year, at distances of order
200 Mpc. If the detectors achieve the so-called
``advanced detector'' sensitivity \cite{LIGO},
then the limiting
distance will be of the order of 1 Gpc, and the event
rate will be increased by a factor of order 100.

The emission of gravitational waves by a compact binary
system causes the orbits to shrink and the orbital
frequency to increase \cite{Thorne1987}.
The gravitational-wave frequency
is given by twice the orbital frequency; it also
increases as the system evolves. The LIGO and
Virgo detectors are designed to operate in the frequency
band between approximately 10 Hz and 1000 Hz \cite{LIGO}.
This corresponds to the last several minutes of inspiral,
and possibly the final coalescence (depending on the
size of the masses involved), of a compact binary
system \cite{Cutleretal}.

During this late stage of orbital evolution the orbital
velocity is large ($v/c$ ranging from
approximately 0.07 to 0.35 if the binary system is that
of two neutron stars \cite{Will}),
and so is the gravitational
field [the dimensionless gravitational potential is of
order $(v/c)^2$]. Accurate modeling of the inspiral must
therefore take general-relativistic effects carefully into
account. On the other hand, other complications, such as tidal
interactions (important only during the last few orbital
cycles \cite{BildstenCutler,Kochanek})
and orbital eccentricity (reduced to very
small values by gravitational radiation reaction
\cite{Peters,LincolnWill}) can be safely ignored.

The gravitational waves emitted by a coalescing compact
binary carry information about the source --- the wave
forms depend on the parameters describing the source. These
include the orientation and position of the source in
the sky, its distance, and the masses and spins
of the companions. A major goal in detecting
these waves is to extract this information
\cite{Thorne1987,Schutz,CutlerFlanagan}.

Because the wave forms can be expressed, at least in
principle, as known functions of the source parameters,
the technique of matched filtering can be used
to estimate the values of these parameters
\cite{Thorne1987,Cutleretal,CutlerFlanagan}: One
cross-correlates the gravitational-wave signal with a
set of theoretical wave forms (templates); the template
with parameters closest to the actual values maximizes
the signal-to-noise ratio. For this
technique to work, the gravitational-wave signal must be
accurately predicted in the full range of frequencies
described above. It can be
estimated that the waves oscillate a number of times of
order $10^4$ as the frequency sweeps from 10 Hz to
1000 Hz \cite{Will}.
To yield a large signal-to-noise ratio the template
must stay in phase with the measured
signal. The accurate prediction of the {\it phasing} of the
wave is therefore a central goal for theorists
\cite{Cutleretal,FinnChernoff}; an accuracy better than
one part in $10^4$ is required.

Gravitational radiation reaction causes $f$, the
gravitational-wave frequency, to increase as the system
evolves. The
phasing of the wave is determined by $\dot{f}(f)$, the
rate of change of frequency as a function of frequency. If
the orbital energy $E$ is known as a function of $f$, then
$\dot{E}(f)$ determines the phasing. Below we will assume
that the relation $E(f)$ is known (or can be calculated),
and focus on the energy loss.

\subsection{Energy loss: post-Newtonian theory}

The energy radiated by a compact binary system can be
calculated, approximately, using post-Newtonian
theory \cite{Will}.
The post-Newtonian approximation is based upon an assumption
about the smallness of the orbital velocity. In units
in which $c=1$ (we also put $G=1$), we demand $v \ll 1$.
On the other hand, nothing is assumed about the mass
ratio: If $M$ denotes total mass and $\mu$ reduced mass,
then $\mu/M$ is allowed to cover the whole range $(0,1/4]$.
For the problem under consideration $v$ is small only
during the early portion of the inspiral, when $f$ is
near 10 Hz. It is therefore expected that calculations
will have to be pushed to extremely high order in
$v$. An intensive effort to do just this is now
underway \cite{Will}.

The phrase ``post-Newtonian theory'' is used here loosely.
A wave-generation formalism suitable for high-order calculations
was developed by Blanchet and Damour
\cite{BlanchetDamour,DamourIyer,Blanchet}. This combines a
post-Newtonian expansion (based, in effect, on the assumption
$c \to \infty$), for the region of spacetime corresponding to
the near zone, with
a post-Minkowskian expansion (based, in effect, on the
assumption $G \to 0$), for the region of spacetime outside
the source. The external metric is matched to the near-zone
metric, and the radiative part of the gravitational field is
extracted. The radiative field is characterized by an infinite
set of multipole moments; the relation between these radiative
moments and the source moments is revealed by matching.

The rate at which energy is radiated can be calculated from
the radiative moments, and expressed as an expansion in powers
of $v$. (In fact, the expansion also involves $\ln v$ at high
order \cite{BlanchetDamour,TagoshiSasaki,TagoshiNakamura}.)
Thus far, the energy loss has been calculated accurately
to order $v^4$, or post$^{2}$-Newtonian order. Schematically,
\begin{equation}
\dot{E} = \dot{E}_{\rm QF} \Bigl[ 1 + O(v^2) +
O(v^3) + O(v^4) + \cdots \Bigr],
\label{1.1}
\end{equation}
where $\dot{E}_{\rm QF} = (32/5)(\mu/M)^2 v^{10}$ is
the leading-order (Newtonian) quadrupole-formula
expression \cite{Thorne1987}.
Here and below, we define $v \equiv (M\Omega)^{1/3} =
(\pi M f)^{1/3}$, where $\Omega$ is the angular velocity.

In Eq.~(\ref{1.1}) the $O(v^2)$ term was first calculated by
Wagoner and Will \cite{WagonerWill}
and is due to post-Newtonian corrections
to the equations of motion and wave generation. The
$O(v^3)$ term comes from two different effects: (i) tail terms
in the waves (the gravitational waves scatter off the
spacetime curvature as they propagate from the near zone
to the far zone), first calculated by Poisson \cite{paperI}
using perturbation theory (see below), and subsequently by
Wiseman \cite{Wiseman} and
Blanchet and Sh\"afer \cite{BlanchetShafer}
using post-Newtonian
theory; and (ii) spin-orbit interactions (present if at least
one of the companions is rotating), first calculated by
Kidder, Will, and Wiseman \cite{KidderWillWiseman}.
The $O(v^4)$ term originates
from spin-spin interactions (if both stars are
rotating), as was calculated by Kidder, Will and Wiseman
\cite{KidderWillWiseman},
and from post$^2$-Newtonian corrections to the equations of
motion and wave generation, as was recently calculated
by Blanchet, Damour, Iyer, Will, and
Wiseman \cite{BDIWW}.
An additional contribution to the $O(v^4)$ term
will be present if at least one of the stars has a
nonvanishing quadrupole moment; this effect has not
yet been calculated, except for the special case
considered by Shibata {\it et al.}~\cite{Shibataetal}
(see below).

\subsection{Energy loss: perturbation theory}

Post-Newtonian theory is based upon a slow-motion,
weak-field approximation. Because the orbital velocity
is large during the late stages of the inspiral, and
because the post-Newtonian expansion is not expected
to converge rapidly \{if at all; the coefficients in
Eq.~(\ref{1.1}) are typically large and grow with
the order \cite{TagoshiSasaki}\},
it is not clear {\it a priori} to which order
in $v$ the calculations must be taken in order to obtain
results of sufficient accuracy. To answer this question
requires a different method of analysis. Perturbation
theory is ideally suited.

In perturbation theory it is assumed that the binary's
mass ratio is very small, $\mu/M \ll 1$, and that the
massive companion is a black hole (rotating or
nonrotating). On the other hand, the orbital velocity
is not restricted: the motion is arbitrarily fast,
and the gravitational field arbitrarily strong. To
calculate, within perturbation theory, the gravitational
waves emitted by such binary systems is the purpose of
this series of papers
\cite{paperI,paperII,paperIII,paperIV,GRR},
as well as that of the work reviewed below.

The perturbation-theory approach can be summarized as
follows.

A particle of mass $\mu$ moves in the gravitational
field of a black hole of mass $M$. (By ``particle'' we mean
an object whose internal structure is of no relevance to the
problem.) The particle possesses a stress-energy tensor
$T^{\alpha \beta}$ which
creates a perturbation in the gravitational field.
If $\mu \ll M$ the Einstein field equations can be solved
perturbatively about the black-hole solution. The
perturbations propagate as gravitational waves. By
solving the perturbation equations one can calculate
the wave forms, as well as the energy and angular momentum
transported. The energy and angular momentum are partly
carried off to infinity, and partly absorbed by the
black hole.

The motion of the particle must be specified before
the perturbation equations are integrated. (The particle's
stress-energy tensor, which acts as a source term in the
equations, is a functional of the world line.)
In general, the motion is affected by the loss of energy
and angular momentum to gravitational waves --- there
is a radiation reaction force \cite{GRR}. However,
the radiation reaction can be neglected
to lowest order in $\mu/M$, and the motion can be taken
to be a geodesic of the black-hole spacetime. The
stress-energy tensor is then completely specified.

Because the source term in the perturbation equations
is known exactly, to integrate these equations amounts to
solving an equation for wave propagation in curved spacetime.
(In post-Newtonian theory, wave propagation is only part
of the problem. One must also solve for the near-zone
physics.) For large velocities the wave equation must be
integrated numerically. When combined with a slow-motion
approximation, the equations of perturbation theory can be
integrated analytically.

\subsection{Perturbation theory: a survey}

The theory of black-hole perturbations (gravitational and
otherwise) has been the topic of a vast literature.
A self-contained summary can be found
in Chandrasekhar's book \cite{Chandra}.
In this series of papers we
have adopted Teukolsky's formulation of
perturbation theory \cite{Teukolsky},
in which the perturbation field
is $\Psi_4$, a complex-valued component of the Weyl tensor.
The equation satisfied by $\Psi_4$ is known as the
Teukolsky equation.

The perturbation formalism was summarized in
paper I \cite{paperI},
and applied to the specific case of circular motion around
a nonrotating black hole. Methods were developed for
integrating the wave equation analytically in the low-frequency
limit, and the energy loss was calculated to order $v^3$ in
the post-Newtonian expansion. These analytical methods were
extended by Sasaki \cite{Sasaki}, and
Tagoshi and Sasaki \cite{TagoshiSasaki} have
obtained $\dot{E}$ to post$^4$-Newtonian order [$O(v^8)$;
see Sec.~V D]. Much was learned from this work about the
structure of the post-Newtonian expansion.

In paper II \cite{paperII}
the perturbation equations were integrated
numerically, and the energy loss was calculated exactly
(apart from numerical error) for a wide range of
orbital frequencies. By fitting a post-Newtonian expansion
to the curve $\dot{E}(f)$ the quality of the approximation
could be determined. It was
concluded that for the purpose of matched filtering,
the expansion must be extended so as to include terms of
order $v^6$ or higher. These numerical calculations
were repeated with much higher accuracy by Tagoshi and
Nakamura \cite{TagoshiNakamura},
who confirmed this conclusion. The analytical
work of Tagoshi and Sasaki \cite{TagoshiSasaki}
is in complete agreement with the numerical work.

The gravitational waves emitted by a particle in eccentric
motion around a Schwarzschild black hole were calculated,
and the orbital evolution under radiation reaction considered,
by Tanaka {\it et al.} \cite{Tanakaetal}.
They discovered that while
radiation reaction always reduces the eccentricity in
slow-motion situations, strong-field radiation reaction
can cause the eccentricity to increase.
This was studied in detail in paper III \cite{paperIII}
for the special case of small eccentricities, and by
Cutler, Kennefick and Poisson \cite{GRR}
(we shall treat this paper
as a honorary member of this series) for the general case.
This increase of eccentricity can be understood as a precursor
effect to the eventual plunging of the orbit, at the end
of the inspiral.

Generalization of the work described above
to the case of a rotating
black hole was the topic of several papers. In the
limit of slow rotation, analytical methods were used in
paper IV \cite{paperIV}
to calculate the corrections to the wave forms and
energy loss due to the black hole's rotation.
These results were generalized by
Shibata {\it et al.}~\cite{Shibataetal}, who
considered a rapidly rotating black hole, as well as
orbits slightly inclined with respect to the hole's
equatorial plane. Both
papers assumed circular orbits. These results were
extended to the case of slightly eccentric orbits
by Tagoshi \cite{Tagoshi}. Numerical results were obtained
by Shibata \cite{Shibata1}, who also considered the case of
eccentric orbits \cite{Shibata2}.

\subsection{Black-hole absorption}

We now turn to a description of the work contained in
this paper.

As mentioned previously, solving the perturbation equations
amounts to solving a wave equation in the background of a
Schwarzschild black hole. Boundary conditions must be
imposed when integrating this equation. The correct
choice is to impose a no-incoming-radiation condition,
which forces the system to lose energy to, and not gain
energy from, gravitational waves.
In practice, this condition is
implemented by imposing outgoing-wave boundary conditions
at infinity, and ingoing-wave boundary conditions at the
black-hole horizon.

In the analytic work of Poisson \cite{paperI},
Sasaki \cite{Sasaki}, and
Tagoshi and Sasaki \cite{TagoshiSasaki},
the horizon boundary conditions were not found to play
a significant role. The black hole could
have been replaced by some regular distribution of matter,
and the horizon boundary conditions replaced by a regularity
condition at the origin (this is the type of condition
which is imposed in post-Newtonian calculations
\cite{BlanchetDamour}), and
the results would have been identical. Due to this
situation, the question of correctly imposing the horizon
boundary conditions was not fully examined.

The apparent irrelevance of the horizon boundary conditions
turns out to be a consequence of two facts. The first is that
only $\dot{E}^\infty$, the energy radiated to infinity, was
calculated in the previous analytical work. The total energy
loss, $\dot{E}$, must also include $\dot{E}^H$, the energy
flowing through the black-hole horizon. The second is that
$\dot{E}^\infty$ was calculated only up to
order $v^8$ in the post-Newtonian expansion.
Sasaki \cite{Sasaki}
has shown [this is essentially a consequence of
Eq.~(\ref{1.4}) below] that the horizon boundary conditions
affect $\dot{E}^\infty$ only at order $v^{18}$ and beyond.

One of the main objectives of this paper is to calculate
$\dot{E}^H$, the rate at which the black hole absorbs
energy, to leading order in a slow-motion approximation.
This calculation is carried out in Sec.~V, using material
derived in preceding sections. We find
\begin{equation}
\dot{E}^H = \dot{E}_{\rm QF} \bigl[v^8 + O(v^{10})\bigr],
\label{1.2}
\end{equation}
so that $\dot{E}^H$ contributes terms of order $v^8$ and
higher to the
post-Newtonian expansion of $\dot{E}$. It can therefore
be said that the black-hole absorption is an effect
occurring at post$^4$-Newtonian order in the energy loss.
This effect utterly dominates that of the horizon boundary
conditions on the energy radiated to infinity.

Nevertheless, the black-hole absorption is a small effect.
In view of the requirement for matched filtering --- energy
loss accurate to post$^3$-Newtonian order
\cite{TagoshiNakamura,paperII} (see subsection D) --- this
effect can be safely ignored. The smallness of
$\dot{E}^H/\dot{E}^\infty$ can be attributed to the
presence of a potential barrier in the vicinity of the
black hole. This barrier is a manifestation of the
spacetime curvature; it
influences the propagation of waves near
the black hole. Since gravitational waves generated by
binary motion have a frequency $f$ such that $\pi M f
= v^3 \ll 1$, most of the waves propagating initially
toward the black hole are reflected off
the potential barrier (see subsection F).
As a result, the black hole is unable to absorb much.

That the black-hole absorption is a small effect is ultimately
due to the extreme degree of compactness of black holes.
How large the absorption would be for a (substantially
less compact) neutron star is unknown.
This question warrants further examination.

\subsection{Integration of the Regge-Wheeler equation}

While integrating the perturbation equations amounts to
solving a wave equation, solving that equation
amounts to integrating the Regge-Wheeler equation
\cite{ReggeWheeler} for
the radial function $X_{\omega\ell}(r)$. Here, $r$
is the Schwarzschild radial coordinate, $\omega$ the
wave's angular frequency, and $\ell$ the spherical-harmonic
index. The Regge-Wheeler equation takes the form
\begin{equation}
\biggl[ \frac{d^2}{dr^{*2}} + \omega^2 + V(r) \biggr]
X_{\omega \ell}(r) = 0,
\label{1.3}
\end{equation}
where $d/dr^* = (1-2M/r) d/dr$ and $V(r) =
(1-2M/r)[\ell(\ell+1)/r^2-6M/r^3]$.
A significant portion of this paper (Sec.~III) is devoted to
developing techniques for integrating Eq.~(\ref{1.3}).
These techniques rely on an assumption of low frequency,
$M\omega \ll 1$, which is appropriate for slow-motion
situations.

Our techniques allow us to calculate the reflection
and transmission coefficients associated with the
potential barrier $V(r)$. Suppose an incident wave
$e^{-i\omega(t+r^*)}$ is sent from $r=\infty$ toward
the black hole. The wave partially reflects off the
potential barrier. The reflected wave
${\cal R}_{\omega\ell} e^{-i\omega(t-r^*)}$ comes
back to $r=\infty$, while the transmitted wave
${\cal T}_{\omega\ell} e^{-i\omega(t+r^*)}$ goes through
the black-hole horizon. In Sec.~III E we find that
the reflection and transmission coefficients are
given by
\begin{equation}
{\cal R}_{\omega\ell} = (\mbox{phase})
\Bigl\{ 1 + O \bigl[|M\omega|^{2(\ell+1)} \bigr] \Bigl\},
\label{1.4}
\end{equation}
and
\begin{eqnarray}
{\cal T}_{\omega\ell} &=&
\frac{2(\ell-2)!(\ell+2)!}{(2\ell)!(2\ell+1)!!}
|2 M\omega|^{\ell+1} (\mbox{phase})
\nonumber \\ & & \mbox{} \times
\Bigl\{
1 + \pi M |\omega| + O\bigl[ (M\omega)^2 \bigr] \Bigr\}.
\label{1.5}
\end{eqnarray}
Expressions for the phase factors, accurate to order
$M\omega$, can be found in Sec.~III E. (In the language
of that section, ${\cal R}_{\omega\ell} =
A^{\rm out}/A^{\rm in}$, and ${\cal T}_{\omega \ell} =
1/A^{\rm in}$.)

Leading-order expressions for the reflection and transmission
coefficients were previously derived by Fackerell
\cite{Fackerell}. (See also Price \cite{Price}
and Thorne \cite{Thorne1972}.) His results
correspond to the $M\omega \to 0$ limit of Eqs.~(\ref{1.4})
and (\ref{1.5}), and therefore omit the $O(M\omega)$ part of
the phase factors and the
$\pi M |\omega|$ correction. These corrections, we should point
out, are not needed for the derivation of Eq.~(\ref{1.2}).

\subsection{Perturbation theory as a wave generation
            formalism}

It is interesting to compare the wave generation
formalism which derives from perturbation theory to the
post-Newtonian formalism of Blanchet and Damour
\cite{BlanchetDamour,DamourIyer,Blanchet}.

The Teukolsky equation can be separated by decomposing the
perturbation field $\Psi_4$ into spin-weighted spherical
harmonics \cite{Goldbergetal}.
{}From the solution at large distances it is
possible to reconstruct the (traceless-transverse)
gravitational-wave field
$h_{ab}^{\rm TT}$, which is then naturally expanded in
tensor spherical harmonics (Sec.~II B).
Apart from an overall factor of $1/r$, the coefficients are
functions of retarded time only, and play the role of
radiative multipole moments. These moments can be expressed
in terms of an integral over the source.
In general the radiative moments are not simply
related to the source moments. However, to leading
order in $v$ in a slow-motion
approximation, the relation {\it is} simple.
This is established in Sec.~IV A.

The situation is analogous in the Blanchet-Damour formalism.
Here, the radiative field is also expanded in multipole
moments, but for convenience a symmetric-tracefree (STF)
representation is favored. (An expansion in STF moments is
entirely equivalent to an expansion in tensor spherical
harmonics \cite{Thorne1980}.)
In general the radiative moments are not simply
related to the source moments, except for the limiting
case of slowly moving sources.

The Blanchet-Damour formalism uses two coordinate systems.
The first, $\{t,\bbox{x}\}$, is rooted to the source, and
chosen so as to make near-zone calculations simple.
However, the true light cones of the spacetime do not
everywhere coincide with the near-zone light cones
(described by $t = |\bbox{x}|$), and
this generates artificial divergences in the radiation
field. To remedy this, Blanchet and Damour
\cite{Blanchet} introduce
radiative coordinates, $\{T,\bbox{X}\}$, which are
adapted to the true light-cone structure of the
spacetime. In terms of these coordinates, the
radiation field is finite. The coordinate systems
are approximately related by $T - |\bbox{X}| = t -
|\bbox{x}| - 2M \ln(|\bbox{x}|/s)$, where
effects nonlinear in the gravitational-wave field
have been neglected. The log term is naturally
interpreted as a Shapiro time delay, and $s$ is an
arbitrary parameter which serves to fix the origin of $T$.

No such coordinate transformation is needed in perturbation
theory. One can work with the usual Schwarzschild coordinates,
and the radiative multipole moments are naturally expressed
as functions of retarded time $u=t-r^*=t-r-2M\ln(r/2M-1)$;
see Sec.~II B. The Shapiro time delay is therefore automatically
incorporated into the retarded time, and the radiative field is
automatically finite.

The Blanchet-Damour formalism has been used to calculate the
tail part of the radiation field \cite{BlanchetDamour}.
Such tails arise because the
curvature of spacetime outside the source scatters the
gravitational waves as they propagate toward the far zone. As
a result, the radiation field at time $T$ does not depend
only on the state of the source at retarded time
$T-|\bbox{X}|$, but also on its entire past history.
This effect modifies both the amplitude and the phase of
the radiative multipole moments.

The tail effect can also be investigated using perturbation
theory, which we do in Sec.~IV B. The results are identical
to that of post-Newtonian theory.

We therefore see that perturbation theory can be used to
recognize and clarify many of the issues that must be
addressed when dealing with post-Newtonian wave generation.
This, we feel, is a useful aspect of our work.

\subsection{Organization of the paper}

We begin in Sec.~II with a brief summary of the perturbation
formalism. This formalism has been manipulated, in the
course of this series of papers, so as to make it as easy
to use as possible. We believe that the present formulation
is as convenient as can possibly be. We will not present
derivations in this section, but refer the reader to
earlier references. Section II A is devoted to a
description of the inhomogeneous Teukolsky equation and
its integration by means of a Green's function. A procedure
to obtain this Green's function in terms of solutions to
the Regge-Wheeler equation is described in schematic terms.
The procedure is detailed in Appendix A. Section II B
describes how to obtain the gravitational-wave field from
the large-distance behavior of the Teukolsky function. The
radiative multipole moments are introduced. Finally,
Sec.~II C gives expressions for the rates at which
the gravitational waves carry energy and angular momentum,
both to infinity and down the black hole.

In Sec.~III we present our techniques for integrating
the Regge-Wheeler equation in the limit of low frequencies.
Two sets of boundary conditions are imposed. The function
$X^H_{\omega\ell}(r)$ satisfies ingoing-wave boundary
conditions at the black-hole horizon, while
$X^\infty_{\omega\ell}(r)$ satisfies outgoing-wave boundary
conditions at infinity \cite{foot1}.
Most of the section is devoted to
$X^H_{\omega\ell}(r)$. Some notation is introduced in Sec.~III A.
The Regge-Wheeler equation is integrated near $r=2M$ in
Sec.~III B, and is integrated in the limit $M\omega \ll 1$
in Sec.~III C. Matching is carried out in Sec.~III D. The
reflection and transmission coefficients are then calculated
in Sec.~III E. Finally, Sec.~III F contains a brief discussion
of the function $X^\infty_{\omega\ell}(r)$.

In Sec.~IV we illustrate how black-hole perturbation theory
can be used as a wave generation formalism. We consider
slowly moving sources, and first (Sec.~IV A) obtain
expressions, valid to leading order in $v$, for the
radiative multipole moments. Then, in Sec.~IV B,
we consider the corrections to the radiative moments which
are due to wave-propagation (tail) effects.

In Sec.~V we present calculations pertaining to
gravitational waves produced by a particle in
circular motion around a black hole. These calculations
are carried out within the slow-motion approximation, to
leading order in $v$. In Secs.~V A and B we derive
expressions for the radiative multipole moments. In Sec.~V C
we calculate the contribution from each moment
to the energy radiated. We consider both the flux at
infinity, and that at the black-hole horizon. Finally,
in Sec.~V D, we consider the role of the black-hole
absorption in the post-Newtonian expansion of the
energy loss.

\section{The perturbation formalism}

The stress-energy tensor associated with a moving particle
perturbs the gravitational field of a nonrotating black hole.
We describe this perturbation using the Teukolsky formalism
\cite{Teukolsky}, which we review below.
The following presentation will be brief; missing
details can be found in Refs.~\cite{paperI,paperIII,GRR},
and in Appendix A.

\subsection{The Teukolsky equation and its solution}

In the Teukolsky formalism, gravitational perturbations of
the Schwarzschild black hole are represented by the complex-valued
function $\Psi_4 = -C_{\alpha\beta\gamma\delta}
n^\alpha \bar{m}^\beta n^\gamma \bar{m}^\delta$. Here,
$C_{\alpha\beta\gamma\delta}$ is the Weyl tensor, and
$n^\alpha = \frac{1}{2} (1,-f,0,0)$, $\bar{m}^\alpha =
(0,0,1,-i\csc\theta)/\sqrt{2} r$
(in the $\{t,r,\theta,\phi\}$ Schwarzschild coordinates)
are members of an orthonormal
null tetrad. We have introduced $f=1-2M/r$, and a bar denotes
complex conjugation.

The Weyl scalar can be decomposed into Fourier-harmonic
components according to
\begin{equation}
\Psi_4 = \frac{1}{r^4} \int_{-\infty}^\infty
d \omega \sum_{\ell m} R_{\omega \ell m} (r)
\mbox{}_{-2} Y_{\ell m} (\theta,\phi) e^{-i \omega t},
\label{2.1}
\end{equation}
where $\mbox{}_{s} Y_{\ell m} (\theta,\phi)$ are spin-weighted
spherical harmonics \cite{Goldbergetal}.
The sums over $\ell$ and $m$ are
restricted to $-\ell \leq m \leq \ell$ and $\ell \geq 2$.
The radial function $R_{\omega\ell m}(r)$ satisfies the
inhomogeneous Teukolsky equation \cite{Teukolsky},
\begin{equation}
\biggl[ r^2 f \frac{d^2}{dr^2} - 2(r - M) \frac{d}{dr}
+ U(r) \biggr] R_{\omega \ell m} (r) = - T_{\omega \ell m} (r),
\label{2.2}
\end{equation}
with
\begin{equation}
U(r) = f^{-1}
\bigl[ (\omega r)^2 - 4 i \omega (r - 3M) \bigr]
-(\ell-1)(\ell+2).
\label{2.3}
\end{equation}

The source term to the right-hand side of
Eq.~(\ref{2.2}) is constructed from the
particle's stress-energy tensor,
\begin{equation}
T^{\alpha\beta}(x) = \mu \int d\tau\,
u^\alpha u^\beta \delta^{(4)}
\bigl[x-x'(\tau)\bigr],
\label{2.4}
\end{equation}
where $x$ represents the spacetime event and $x'(\tau)$
the particle's world line with tangent vector
$u^\alpha = dx'^\alpha/d\tau$ ($\tau$ denotes
proper time). The first step is to obtain the projections
$\mbox{}_0 T = T_{\alpha\beta} n^\alpha n^\beta$,
$\mbox{}_{-1} T = T_{\alpha\beta} n^\alpha \bar{m}^\beta$, and
$\mbox{}_{-2} T = T_{\alpha\beta} \bar{m}^\alpha \bar{m}^\beta$.
Then one calculates the Fourier-harmonic components
$\mbox{}_s T_{\omega \ell m} (r)$ according to
\begin{equation}
\mbox{}_s T_{\omega \ell m} (r) = \frac{1}{2\pi}
\int dt\, d\Omega\, \mbox{}_s T\, \mbox{}_s \bar{Y}_{\ell m}
(\theta,\phi) e^{i \omega t},
\label{2.5}
\end{equation}
where $d\Omega$ is the element of solid angle. Finally,
$T_{\omega \ell m}(r)$ is obtained by applying a certain
differential operator to each
$\mbox{}_s T_{\omega \ell m}(r)$, and then summing over
$s$ \cite{paperI,SasakiNakamura}.
Schematically, $T = \sum_s \mbox{}_s D \mbox{}_s T$,
where $\mbox{}_s D$ (other indices suppressed for
notational simplicity) are the differential operators.
(See Appendix A for details.)

The inhomogeneous Teukolsky equation (\ref{2.2}) can be
integrated by means of a Green's function
\cite{Detweiler}. The Green's
function is constructed from two linearly independent
solutions to the {\it homogeneous} equation, so that
$\Psi_4$ satisfies a no-incoming-radiation condition.
Schematically (indices suppressed), we have $G(r,r') =
R^H(r_<) R^\infty(r_>)$, where $G(r,r')$ is the Green's
function, $R^H(r)$ the solution to the homogeneous
equation satisfying ingoing-wave boundary conditions
at the black-hole horizon, $R^\infty(r)$ the solution
satisfying outgoing-wave boundary conditions at infinity,
and $r_<$ ($r_>$) denotes the lesser (greater) of $r$ and
$r'$. Schematically also, the solution takes the form $R(r) =
\int dr'\, G(r,r') T(r') = \sum_s \int dr'\, G(r,r') \mbox{}_s
D \mbox{}_s T(r')$.

It is then useful to define the adjoint
operators $\mbox{}_s D^\dagger$ so that $R(r)$ can be more
conveniently expressed as $R(r) = \sum_s \int dr'\, \mbox{}_s
T(r') \mbox{}_s D^\dagger G(r,r')$. The last step
consists of invoking the Chandrasekhar transformation
\cite{ChandraTransf},
which relates solutions to the homogeneous Teukolsky
equation to that of the Regge-Wheeler equation
\cite{ReggeWheeler}. Thus,
$G(r,r')$ can be conveniently written in terms of
linearly independent solutions of the Regge-Wheeler
equation. (Relevant details can be found in Appendix A.)

The Regge-Wheeler equation \cite{ReggeWheeler}
takes the form
\begin{equation}
\biggl[ \frac{d^2}{dr^{*2}} + \omega^2 - V(r) \biggr]
X_{\omega \ell}(r) = 0,
\label{2.6}
\end{equation}
where $d/dr^* = f d/dr$ and $V(r) = f[\ell(\ell-1)/r^2
-6M/r^3]$. The solutions of interest are
$X^H_{\omega \ell}(r)$, which satisfies ingoing-wave
boundary conditions at the black-hole horizon, and
$X^\infty_{\omega\ell}(r)$, which satisfies outgoing-wave
boundary conditions at infinity \cite{foot1}.
More precisely, these
functions are defined so as to have the following
asymptotic behaviors:
\begin{eqnarray}
X^H_{\omega\ell}(r\to 2M) &\sim& e^{-i \omega r^*},
\nonumber \\
X^H_{\omega\ell}(r\to\infty) &\sim&
A^{\rm in}_{\omega\ell} e^{-i \omega r^*} +
A^{\rm out}_{\omega\ell} e^{i \omega r^*},
\label{2.7} \\
X^\infty_{\omega\ell}(r\to\infty) &\sim&
e^{i \omega r^*}. \nonumber
\end{eqnarray}
Here, $r^* = r + 2M \ln (r/2M - 1)$, and
$A^{\rm in}_{\omega\ell}$ and
$A^{\rm out}_{\omega\ell}$ are constants.
It follows from the conservation of the Wronskian that
$X^\infty_{\omega\ell}(r\to 2M) \sim
-\bar{A}^{\rm out}_{\omega\ell} e^{-i \omega r^*} +
A^{\rm in}_{\omega\ell} e^{i \omega r^*}$.

Schematically, the Chandrasekhar transformation
\cite{ChandraTransf} takes the
form $R^{H,\infty} = C X^{H,\infty}$,
where $C$ is a second-order differential operator.
Thus, the effective Green's functions $\mbox{}_s D^\dagger
G(r,r')$ can be expressed in terms of
$X^{H,\infty}(r)$
and their derivatives. Because the Regge-Wheeler functions
satisfy a second-order differential equation, the final
expressions involve the functions and their first derivatives
only. (See Appendix A for details.)

At large radii, the solution to the inhomogeneous
Teukolsky equation is given by \cite{paperIII}
\begin{equation}
R_{\omega\ell m} (r\to\infty) \sim
\mu \omega^2 Z^H_{\omega \ell m} r^3 e^{i \omega r^*}.
\label{2.8}
\end{equation}
Near the black-hole horizon \cite{paperIII},
\begin{equation}
R_{\omega\ell m} (r\to 2M) \sim
\mu \omega^3 Z^\infty_{\omega \ell m}
r^4 f^2 e^{-i \omega r^*}.
\label{2.9}
\end{equation}
After carrying out the manipulations described above,
the ``amplitudes'' $Z^{H,\infty}_{\omega \ell m}$ are
found to be given by
\begin{eqnarray}
Z^{H,\infty}_{\omega \ell m} &=&
i \pi \Bigl[\mu \omega \kappa^{H,\infty}_{\omega \ell}
A^{\rm in}_{\omega \ell}\Bigr]^{-1} \sum_s \mbox{}_s p_\ell
\nonumber \\
& & \mbox{} \times \int_{2M}^\infty dr\,
r f^{-2} \mbox{}_s T_{\omega \ell m} (r)\,
\mbox{}_s \Gamma_{\omega \ell}
X^{H,\infty}_{\omega \ell} (r).
\label{2.10}
\end{eqnarray}
We have introduced the following symbols:
\begin{eqnarray}
\kappa^H_{\omega \ell} &=&
{\textstyle \frac{1}{4}} \bigl[
(\ell-1)\ell(\ell+1)(\ell+2) - 12 i M \omega \bigr],
\nonumber \\
& & \label{2.11} \\
\kappa^\infty_{\omega \ell} &=& -16 (1-2iM\omega)(1-4iM\omega)
(M\omega)^3, \nonumber
\end{eqnarray}
and
\begin{equation}
\mbox{}_s p_\ell = \left\{
\begin{array}{ll}
2 \bigl[(\ell-1)\ell(\ell+1)(\ell+2) \bigr]^{1/2} & s=0 \\
2 \bigl[2(\ell-1)(\ell+2)\bigr]^{1/2} & s=-1 \\
1 & s=-2
\end{array} \right. .
\label{2.12}
\end{equation}
We also have
\begin{eqnarray}
\mbox{}_0 \Gamma_{\omega \ell} &=&
2(1-3M/r+i\omega r) r f \frac{d}{dr}
+ f \bigl[ \ell(\ell+1) - 6M/r \bigr]
\nonumber \\ & & \mbox{} +
2i\omega r (1 - 3M/r + i\omega r), \nonumber \\
\mbox{}_{-1} \Gamma_{\omega \ell} &=& -f \biggl\{
\bigl[ \ell(\ell+1)+2i\omega r \bigr] r f \frac{d}{dr}
\nonumber \\ & & \mbox{}
+ \ell(\ell+1)(f+i\omega r) - 2 (\omega r)^2 \biggr\},
\label{2.13} \\
\mbox{}_{-2} \Gamma_{\omega\ell} &=& f^2 \biggl\{
2 \bigl[ (\ell-1)(\ell+2)+6M/r \bigr] rf \frac{d}{dr}
\nonumber \\ & & \mbox{}
+ (\ell-1)(\ell+2)\bigl[\ell(\ell+1) + 2i\omega r \bigr]
+ 12 f M/r \biggr\}. \nonumber
\end{eqnarray}
As noted previously, $\mbox{}_s \Gamma_{\omega\ell}$
are {\it first-order} differential operators. That this
is so is convenient both for analytical and numerical
evaluation of $Z^{H,\infty}_{\omega\ell m}$.

\subsection{Waveform and multipole moments}

The gravitational-wave field can be obtained from the
asymptotic behavior of $\Psi_4$ at large distances.
Choosing the $\theta$ and $\phi$ directions as
polarization axes, the two fundamental polarizations
of the gravitational waves can be expressed as
\cite{paperIII}
\begin{equation}
h \equiv h_+ - i h_\times =
\frac{2\mu}{r} \sum_{\ell m} \tilde{Z}^H_{\ell m} (u)
\mbox{}_{-2} Y_{\ell m}(\theta,\phi),
\label{2.14}
\end{equation}
where $u=t-r^*$, and
\begin{equation}
\tilde{Z}^H_{\ell m} (u) = \int_{-\infty}^\infty
d\omega \, Z^H_{\omega \ell m} e^{-i \omega u}.
\label{2.15}
\end{equation}
The traceless-transverse gravitational-wave tensor is
then
\begin{equation}
h^{\rm TT}_{ab} = h m_a m_b +
\bar{h} \bar{m}_a \bar{m}_b,
\label{2.16}
\end{equation}
where latin indices denote spatial components.

It is clear that the $\tilde{Z}^H_{\ell m}(u)$ represent
the radiative multipole moments of the gravitational
field. To relate these quantities to the commonly used
definitions, we substitute Eq.~(\ref{2.14}) into (\ref{2.16})
and rewrite in terms of the Mathews tensor spherical
harmonics \cite{Thorne1980}.
The result is Eq.~(4.3) of Ref.~\cite{Thorne1980},
\begin{equation}
h^{\rm TT}_{ab} = \frac{1}{r} \sum_{\ell m}
\Bigl[ {\cal I}_{\ell m} (u) T^{E2,\ell m}_{ab} +
{\cal S}_{\ell m} (u) T^{B2,\ell m}_{ab} \Bigr],
\label{2.17}
\end{equation}
where the $T$-tensors are the spherical harmonics. We thus find
that the mass multiple moments are given by
\begin{equation}
{\cal I}_{\ell m}(u) = \sqrt{2} \mu
\Bigl[ \tilde{Z}^H_{\ell m}(u) + (-1)^m
\overline{\tilde{Z}^H}_{\ell,-m}(u) \Bigr],
\label{2.18}
\end{equation}
while the current multipole moments are given by
\begin{equation}
{\cal S}_{\ell m}(u) = \sqrt{2} i \mu
\Bigl[ \tilde{Z}^H_{\ell m}(u) - (-1)^m
\overline{\tilde{Z}^H}_{\ell,-m}(u) \Bigr].
\label{2.19}
\end{equation}

When the source is confined to the equatorial plane,
the identity $\bar{Z}^H_{-\omega,\ell,-m} = (-1)^\ell
Z^H_{\omega\ell m}$ holds (see paper III \cite{paperIII}
for a derivation). This simplifies the expressions for
the mass and current multipole moments.

\subsection{Energy and angular momentum fluxes}

Equations (\ref{2.8}) and (\ref{2.9}) can be used to
calculate the rates at which energy and angular momentum
are radiated to infinity and absorbed by the black hole.
Here we consider only the special case of circular orbits;
more general results can be found in Ref.~\cite{paperIII}.

It will be shown in Sec.~V that for circular orbits, the
frequency spectrum of the gravitational waves contains
all the harmonics of the orbital frequency $\Omega \equiv
d\phi/dt$, so that
\begin{equation}
Z^{H,\infty}_{\omega \ell m} = Z^{H,\infty}_{\ell m}
\delta(\omega - m\Omega).
\label{2.20}
\end{equation}
For this special case, the energy flux at infinity is
given by \cite{TeukolskyPress}
\begin{equation}
\dot{E}^\infty = \frac{1}{4\pi} \Bigl(\frac{\mu}{M}\Bigr)^2
\sum_{\ell m} (M\omega)^2 |Z^H_{\ell m}|^2,
\label{2.21}
\end{equation}
while the energy flux at the event horizon is
\cite{TeukolskyPress}
\begin{equation}
\dot{E}^H = \frac{1}{4\pi} \Bigl( \frac{\mu}{M} \Bigr)^2
\sum_{\ell m} \alpha_{\ell m} |Z^\infty_{\ell m}|^2,
\label{2.22}
\end{equation}
where
\begin{equation}
\alpha_{\ell m} = \frac{2^{12} \bigl[1+4(M\omega)^2\bigr]
\bigl[1+16(M\omega)^2]}{\bigl| (\ell-1)\ell(\ell+1)(\ell+2)
- 12iM\omega \bigr|^2}\, (M\omega)^8.
\label{2.23}
\end{equation}
In Eqs.~(\ref{2.21}) and (\ref{2.22}), a dot denotes differentiation
with respect to coordinate time $t$, and $\omega \equiv m \Omega$. The
total amount of energy radiated by the gravitational waves is
$\dot{E}=\dot{E}^\infty + \dot{E}^H$.

The angular momentum fluxes can be obtained from the relations
$\dot{E}^\infty = \Omega \dot{L}^\infty$ and
$\dot{E}^H = \Omega \dot{L}^H$, valid for any radiating system
in rigid rotation \cite{ZeldovichNovikov}.

\section{Integration of the Regge-Wheeler equation}

Our objective in this section is to integrate
Eq.~(\ref{2.6}) in the limit $M\omega \ll 1$.
Techniques for solving this problem were developed by
Poisson in the first paper in this series \cite{paperI},
and then extended by Sasaki \cite{Sasaki}. We extend these
techniques further here, by explicitly considering
the issue of the boundary conditions at the black-hole
horizon.

Most of this section will be devoted to the
``ingoing-wave'' Regge-Wheeler function
$X^H_{\omega\ell}(r)$. The ``outgoing-wave''
function $X^\infty_{\omega\ell}(r)$ will
be briefly considered in subsection F. For
notational simplicity we will, throughout this
section, suppress the use of the $\omega\ell$
suffix.

\subsection{Preliminaries}

For concreteness we assume that the wave
frequency $\omega$ is positive. Results for
negative frequencies must be derived
separately \cite{foot}. For convenience we
define
\begin{equation}
z = \omega r, \quad \varepsilon = 2M\omega, \quad
z^* = z + \varepsilon \ln(z-\varepsilon).
\label{3.1}
\end{equation}
Both $z$ and $\varepsilon$ are dimensionless, and we
note that $z^*=\omega r^*+ \varepsilon \ln \varepsilon$;
$r^*$ was introduced in Eq.~(\ref{2.7}).

With these definitions the Regge-Wheeler equation becomes
\begin{equation}
\biggl\{ \frac{d^2}{dz^{*2}} + 1 - f
\biggl[ \frac{\ell(\ell+1)}{z^2} -
\frac{3\varepsilon}{z^3} \biggr] \biggr\}
X(z) = 0,
\label{3.2}
\end{equation}
where $f=1-\varepsilon/z$. The ingoing-wave
Regge-Wheeler function
satisfies the following boundary conditions:
\begin{eqnarray}
X^H(z\to\varepsilon) &\sim& e^{i\varepsilon \ln \varepsilon}
e^{-i z^*}, \nonumber \\
& & \label{3.3} \\
X^H(z\to\infty) &\sim& A^{\rm in}
e^{i\varepsilon \ln \varepsilon} e^{-iz^*}
+ A^{\rm out} e^{-i\varepsilon \ln \varepsilon} e^{iz^*}.
\nonumber
\end{eqnarray}

To carry out the integration of the Regge-Wheeler equation it
is useful to follow Sasaki \cite{Sasaki}
and introduce an auxiliary function
$Y^H(z)$ defined by
\begin{equation}
Y^H(z) = z^{-1} e^{i\varepsilon \ln(z-\varepsilon)}
X^H(z).
\label{3.4}
\end{equation}
The great advantage of dealing with $Y^H(z)$ instead of
$X^H(z)$ directly will be become apparent below. The
auxiliary function satisfies
\begin{eqnarray}
\biggl\{ z(z-\varepsilon) \frac{d^2}{dz^2} +
\bigl[ 2(1-i\varepsilon)z-\varepsilon \bigr] \frac{d}{dz}
\hspace{1.2cm} & & \nonumber \\ \mbox{} +
\Bigl[z^2 - \ell(\ell+1) + \frac{\varepsilon}{z}
(4-iz+z^2) \Bigr] \biggr\} Y^H(z) &=& 0.
\label{3.5}
\end{eqnarray}
Equations (\ref{3.3}) and (\ref{3.4}) further imply
\begin{equation}
Y^H(z\to\varepsilon) \sim \varepsilon^{-1}
e^{i \varepsilon (\ln \varepsilon - 1)}
\label{3.6}
\end{equation}
and
\begin{eqnarray}
Y^H(z\to\infty) &\sim&
A^{\rm in} e^{i\varepsilon\ln\varepsilon} z^{-1} e^{-iz}
\nonumber \\ & & \mbox{} +
A^{\rm out} e^{-i\varepsilon\ln\varepsilon} z^{-1}
e^{i(z+ 2\varepsilon \ln z)}.
\label{3.7}
\end{eqnarray}

For future reference we shall now rewrite Eq.~(\ref{3.7}) in
a different form. We first expand the right-hand side in powers
of $\varepsilon$, which below will treated as a small number.
(This assumption has not yet been used.) We also invoke the
spherical Bessel functions $j_\ell(z)$ and $n_\ell(z)$, which
will play a prominent role below, via the asymptotic relations
$z^{-1} e^{\pm i z} \sim (\pm i)^{\ell+1} [j_\ell(z\to\infty)
\pm i n_\ell(z\to\infty)]$. We thus obtain
\begin{eqnarray}
Y^H(z\to\infty) &\sim& \bigl[a_+
+ 2 i \tilde{a} \varepsilon \ln z
+ O(\varepsilon^2) \bigr] j_\ell(z\to\infty) \nonumber \\
& & \mbox{} -
\varepsilon \bigl[i a_- + 2 \tilde{a} \ln z
+ O(\varepsilon) \bigr]
n_\ell(z\to\infty). \nonumber \\ & &
\label{3.8}
\end{eqnarray}
In terms of $a_\pm$ we have
\begin{eqnarray}
A^{\rm in} &=& {\textstyle \frac{1}{2}} (i)^{\ell+1}
(a_+ + \varepsilon a_-) e^{-i\varepsilon\ln\varepsilon},
\nonumber \\
A^{\rm out} &=& {\textstyle \frac{1}{2}} (-i)^{\ell+1}
(a_+ - \varepsilon a_-) e^{i\varepsilon\ln\varepsilon},
\label{3.9} \\
\tilde{a} &=& {\textstyle \frac{1}{2}}
(a_+ - \varepsilon a_-).
\nonumber
\end{eqnarray}

\subsection{Auxiliary function: solution for $z\ll 1$}

We first integrate Eq.~(\ref{3.5}) in the limit
$z \ll 1$. We shall also treat $\varepsilon$ as a small
number, but leave arbitrary the ratio $z/\varepsilon$.
Accordingly, we neglect the $z^2$ terms within the
large square brackets of Eq.~(\ref{3.5}): the first one
can be neglected in front of $\ell(\ell+1)$, while the
second is negligible in front of $4$. All other terms are
kept in the equation.

The resulting equation can easily be solved if we change the
independent variable to $x=1-z/\varepsilon$ and the dependent
variable to $Z = (\varepsilon/z)^2 Y$. We then obtain
\begin{eqnarray}
x(x-1)Z'' + \bigl[2(3-i\varepsilon)x-(1-2i\varepsilon)\bigr] Z'
& & \nonumber \\ \mbox{} +
\bigl[6-\ell(\ell+1)-5i\varepsilon\bigr] Z &=& 0,
\label{3.10}
\end{eqnarray}
where a prime denotes differentiation with respect to $x$. This
is the hypergeometric equation, with parameters
\begin{eqnarray}
a &=& -(\ell-2) - i \varepsilon + O(\varepsilon^2), \nonumber \\
b &=& \ell+3-i \varepsilon + O(\varepsilon^2), \label{3.11} \\
c &=& 1-2i\varepsilon. \nonumber
\end{eqnarray}
The two linearly independent solutions are $F(a,b;c;x)$ and
$x^{1-c} F(a+1-c,b+1-c;2-c;x)$, where $F$ is the hypergeometric
function. However, only the first solution
is regular at $x=0$ ($z=\varepsilon$). Compatibility with
Eq.~(\ref{3.6}) therefore implies that the second solution must
be rejected. We therefore obtain
\begin{equation}
Y^H(z\ll 1) = \varepsilon^{-1}
e^{i \varepsilon (\ln\varepsilon - 1)} (z/\varepsilon)^2
F(a,b;c;1-z/\varepsilon).
\label{3.12}
\end{equation}
Equation (\ref{3.12}) enforces the correct boundary condition
at the black-hole horizon.

For later use we now evaluate $Y^H(z)$ in the limit
$\varepsilon \ll z \ll 1$. First, a straightforward calculation,
using Eq.~(15.3.7) of Abramowitz and Stegun
\cite{AbramowitzStegun}, reduces Eq.~(\ref{3.12}) to
\begin{eqnarray}
Y^H(\varepsilon \ll z \ll 1) &=& \varepsilon^{-1}
e^{i\varepsilon(\ln\varepsilon-1)}
\frac{\Gamma(c) \Gamma(b-a)}{\Gamma(b)\Gamma(c-a)}
\Bigl( \frac{z}{\varepsilon} \Bigr)^{\ell+i\varepsilon}
\nonumber \\ & & \mbox{} \times \biggl\{
1 - \biggl[ \frac{(\ell-2)(\ell+2)}{2\ell} + i\varepsilon \biggr]
\frac{\varepsilon}{z} \nonumber \\ & & \mbox{} +
O\bigl[ (\varepsilon/z)^2 \bigr] \biggr\}.
\label{3.13}
\end{eqnarray}
Next, we expand the complex exponential, and the $\Gamma$
functions according to $\Gamma(n+\delta) =
(n-1)![1 + \psi(n) \delta + O(\delta^2)]$, in powers of
$\varepsilon$. [Here, $n$ is an integer and $\psi(n)=-\gamma
+\sum_{k=1}^{n-1} k^{-1}$, where $\gamma \simeq 0.57721$ is
the Euler constant, is the digamma function.] Finally, we
obtain
\begin{eqnarray}
Y^H(\varepsilon \ll z \ll 1) &=&
\frac{(2\ell)!}{(\ell-2)!(\ell+2)!}
\frac{z^\ell}{\varepsilon^{\ell+1}}
\biggl[1 + i \varepsilon (\alpha_\ell +\ln z)
\nonumber \\ & & \mbox{} -
\frac{(\ell-2)(\ell+2)}{2\ell} \frac{\varepsilon}{z} +
O(\varepsilon^2) \biggr].
\label{3.14}
\end{eqnarray}
We have defined
\begin{equation}
\alpha_\ell = 2\gamma + \psi(\ell-1) + \psi(\ell+3) - 1.
\label{3.15}
\end{equation}

\subsection{Auxiliary function: solution for
            $\varepsilon \ll 1$}

We now integrate Eq.~(\ref{3.5}) in the limit $\varepsilon \ll 1$.
It will also be assumed that $z \gg \varepsilon$, but $z$ is no
longer restricted to be much smaller than unity.

We first rewrite Eq.~(\ref{3.5}) by sending to the right-hand side
all terms which are linear in $\varepsilon$; the left-hand side is
then independent of $\varepsilon$. To solve this equation we
proceed iteratively \cite{Sasaki}, by
setting
\begin{equation}
Y^H(z) = \sum_{n=0}^{\infty} \varepsilon^n Y^{H(n)}(z).
\label{3.16}
\end{equation}
A short calculation then shows that each $Y^{H(n)}(z)$ satisfies
\begin{equation}
\biggl[ z^2 \frac{d^2}{dz^2} + 2z \frac{d}{dz} + z^2 -
\ell(\ell+1) \biggr] Y^{H(n)}(z) = S^{H(n)}(z),
\label{3.17}
\end{equation}
where
\begin{eqnarray}
S^{H(n)}(z) &=& \frac{1}{z} \biggl[ z^2 \frac{d^2}{dz^2} +
z(1+2iz) \frac{d}{dz} \nonumber \\
& & \mbox{} - (4-iz+z^2) \biggr] Y^{H(n-1)}(z).
\label{3.18}
\end{eqnarray}
Equation (\ref{3.17}) is an inhomogeneous spherical Bessel
equation. It is the simplicity of this equation which motivated
the introduction of the auxiliary function \cite{Sasaki}.

The zeroth-order solution, $Y^{H(0)}(z)$, satisfies the homogeneous
spherical Bessel equation, and must therefore be a linear combination
of $j_\ell(z)$ and $n_\ell(z)$. Compatibility with Eq.~(\ref{3.14})
then demands
\begin{equation}
Y^{H(0)}(z) = B j_\ell(z),
\label{3.19}
\end{equation}
where $B$ is constant to be determined.

The procedure to
obtain $Y^{H(1)}(z)$ was described in detail in
Ref.~\cite{Sasaki}.
In short, Eq.~(\ref{3.19}) is substituted into Eq.~(\ref{3.18})
to obtain $S^{H(1)}$, and Eq.~(\ref{3.17}) is integrated using the
Green's function $G(z,z') = j_\ell(z_<) n_\ell(z_>)$. The resulting
integrals can be evaluated explicitly, and the final result is
\cite{Sasaki}
\begin{eqnarray}
Y^{H(1)}(z)/B &=& \Bigl(a^{(1)} + i \ln z - \mbox{Si} 2z
\Bigr) j_\ell \nonumber \\ & & \mbox{}
+ \Bigl(b^{(1)} - \ln 2z + \mbox{Ci} 2z
\Bigr) n_\ell \nonumber \\
& & \mbox{} + z^2(n_\ell j_0 - j_\ell n_0) j_0 \nonumber \\
& & \mbox{} +
\sum_{p=1}^{\ell-2} \biggl(\frac{1}{p} + \frac{1}{p+1}\biggr)
z^2 (n_\ell j_p - j_\ell n_p) j_p \nonumber \\
& & \mbox{} - \biggl[ \frac{(\ell-2)(\ell+2)}{2\ell(2\ell+1)}
+ \frac{2\ell-1}{\ell(\ell-1)} \biggr] j_{\ell-1}
\nonumber \\ & & \mbox{} +
\frac{(\ell-1)(\ell+3)}{2(\ell+1)(2\ell+1)} j_{\ell+1}.
\label{3.20}
\end{eqnarray}
Here, $a^{(1)}$ and $b^{(1)}$ are constants of integration
to be determined, $\mbox{Si}x = \int_0^x dt\, t^{-1} \sin t$
is the sine integral, and $\mbox{Ci}x = \gamma + \ln x +
\int_0^x dt\, t^{-1} (\cos t -1)$ is the cosine integral.

\subsection{Matching}

The constants $B$, $a^{(1)}$ and $b^{(1)}$ can be determined
by matching $Y^H(z) = Y^{H(0)}(z) + \varepsilon Y^{H(1)}(z) +
O(\varepsilon^2)$ to Eq.~(\ref{3.14}). This involves the
evaluation of Eqs.~(\ref{3.19}) and (\ref{3.20}) in the
limit $z\ll 1$. This calculation is straightforward, since
the limiting expressions for the spherical Bessel functions,
as well as that for
the sine and cosine integrals, are well known. We
find that $b^{(1)} = -\gamma$ and that $a^{(1)}$ is
arbitrary --- it can be absorbed into the
$O(\varepsilon)$ part of $B$.
Without loss of generality we may set $a^{(1)}=0$.

We note in passing that consistency between Eqs.~(\ref{3.14}),
(\ref{3.19}) and (\ref{3.20}) demands that the following
equation hold:
\begin{eqnarray}
\sum_{n=2}^{\infty} (-1)^n \frac{n-1}{2n(2n)!} (2z)^{2n}
- \sum_{p=1}^{\ell-2} \biggl( \frac{1}{p} + \frac{1}{p+1} \biggr)
z^2 {j_p}^2 & & \nonumber \\
= \frac{z^{2l}\bigl[ 1 + O(z^2) \bigr]}{
(2\ell-1)!!(2\ell-3)!!\ell(\ell-1)}. & &
\label{3.21}
\end{eqnarray}
We have not been able to directly establish the validity of
Eq.~(\ref{3.21}), but have checked that it does indeed hold
for several special cases.

Finally, matching yields the value of $B$, which is
\begin{equation}
B = \frac{(2\ell)!(2\ell+1)!!}{(\ell-2)!(\ell+2)!}
\frac{1}{\varepsilon^{\ell+1}}
\bigl[ 1 + i \varepsilon \alpha_\ell + O(\varepsilon^2) \bigr],
\label{3.22}
\end{equation}
where $\alpha_\ell$ was introduced in Eq.~(\ref{3.15}). The
integration of Eq.~(\ref{3.5}), accurately to first order in
$\varepsilon$ and with boundary condition (\ref{3.6}), is now
completed. In Ref.~\cite{Sasaki}, Sasaki has indicated how to
proceed to higher order in $\varepsilon$.

\subsection{Calculation of $A^{\rm in}$ and $A^{\rm out}$}

We are now in a position to evaluate $Y^H(z)$ in the limit
$z\to\infty$ and to compare with Eq.~(\ref{3.8}) so as to
obtain expressions for the amplitudes $A^{\rm in}$ and
$A^{\rm out}$. Again the calculation is straightforward,
since the asymptotic expressions for the spherical Bessel
functions (as well as that for the sine and cosine integrals)
are well known.

We use such expressions as $j_{\ell-1} \sim -j_{\ell+1} \sim
-n_\ell$ and $z^2 (n_\ell j_p - j_\ell n_p) j_p \sim
\frac{1}{2}[1-(-1)^{\ell-p}]n_\ell$ to calculate
\begin{eqnarray}
Y^H(z\to\infty) &\sim& B \Bigl\{
\bigl[ 1 + \varepsilon( i\ln z - \pi/2 ) \bigr] j_\ell(z\to\infty)
\nonumber \\ & & \mbox{}
- \varepsilon (\ln 2z - \beta_\ell) n_\ell(z\to\infty)
\nonumber \\
& & \mbox{} + O(\varepsilon^2) \Bigr\},
\label{3.23}
\end{eqnarray}
where
\begin{equation}
\beta_\ell = \frac{1}{2} \biggl[
\psi(\ell) + \psi(\ell+1) +
\frac{(\ell-1)(\ell+3)}{\ell(\ell+1)} \biggr].
\label{3.24}
\end{equation}
Comparison with Eqs.~(\ref{3.8}) and (\ref{3.9}) finally
yields
\begin{eqnarray}
A^{\rm in} &=&
\frac{(2\ell)!(2\ell+1)!!}{2(\ell-2)!(\ell+2)!}
\biggl( \frac{i}{\varepsilon} \biggr)^{\ell+1}
e^{-i\varepsilon(\ln 2\varepsilon -\alpha_\ell - \beta_\ell)}
\nonumber \\ & & \mbox{} \times
\biggl[ 1 - \frac{\pi}{2} \varepsilon + O(\varepsilon^2) \biggr],
\nonumber \\
& & \label{3.25} \\
A^{\rm out} &=&
\frac{(2\ell)!(2\ell+1)!!}{2(\ell-2)!(\ell+2)!}
\biggl( \frac{-i}{\varepsilon} \biggr)^{\ell+1}
e^{i\varepsilon(\ln 2\varepsilon +\alpha_\ell - \beta_\ell)}
\nonumber \\ & & \mbox{} \times
\biggl[ 1 - \frac{\pi}{2} \varepsilon + O(\varepsilon^2) \biggr].
\nonumber
\end{eqnarray}

Analytical expressions for $A^{\rm in}$ and $A^{\rm out}$
in the low-frequency regime were
first obtained by Fackerell \cite{Fackerell};
his results correspond to the
$\varepsilon \to 0$ limit of Eq.~(\ref{3.25}). That these
quantities scale like $\varepsilon^{-(\ell+1)}$ in that regime
was first discovered by Price \cite{Price} and
Thorne \cite{Thorne1972}. The $O(\varepsilon)$
corrections to Fackerell's results, we believe, are presented here
for the first time.

It is important to notice that $A^{\rm out}$ is not quite the
complex conjugate of $A^{\rm in}$. (The physical relevance of
the phase factor will be discussed in Sec.~IV B.)
In magnitude, Eq.~(\ref{3.25})
implies that they are equal, $|A^{\rm out}|=|A^{\rm in}|[1 +
O(\varepsilon^2)]$, up to a fractional accuracy of order
$\varepsilon^2$. In fact, a much stronger result follows from
the Wronskian relation $|A^{\rm in}|^2-|A^{\rm out}|^2=1$.
Simple manipulations and use of Eq.~(\ref{3.25}) indeed reveal that
$|A^{\rm out}|$ and $|A^{\rm out}|$ are equal up to a fractional
accuracy of order $\varepsilon^{2(\ell+1)}$.

\subsection{The outgoing-wave Regge-Wheeler function}

We now integrate the Regge-Wheeler equation for
$X^\infty_{\omega\ell}(r)$, the outgoing-wave function.
The method is entirely analogous to that described in
the preceding subsections. We briefly sketch
the most important steps.

We first introduce an auxiliary function $Y^\infty(z)$
defined by
\begin{equation}
Y^\infty(z) = z^{-1} e^{-i\varepsilon \ln(z-\varepsilon)}
X^\infty(z).
\label{3.26}
\end{equation}
{}From Eq.~(\ref{2.7}) we see that the auxiliary function
must satisfy the boundary condition
\begin{equation}
Y^\infty(z\to\infty) \sim e^{-i\varepsilon\ln\varepsilon}
z^{-1} e^{iz}.
\label{3.27}
\end{equation}
It is immediate that $\bar{Y}^\infty(z)$ satisfies the
same differential equation, Eq.~(\ref{3.5}), as $Y^H(z)$.
Since the boundary conditions are imposed at $z=\infty$, we
shall not need to solve Eq.~(\ref{3.5}) in the limit $z\ll 1$
(as described in subsection B).
Instead we can proceed with an analysis
similar to that contained in subsection C.

The zeroth-order solution $Y^{\infty(0)}$ must
be identified, up to a normalization constant,  with the
spherical Hankel function $h^{(1)}_\ell(z)$, whose asymptotic
behavior is identical to Eq.~(\ref{3.27}). We therefore
have
\begin{equation}
Y^{\infty(0)}(z) = C h^{(1)}_\ell (z),
\label{3.28}
\end{equation}
where $C$ is a constant to be determined. Taking the
complex conjugate, we get $\bar{Y}^{\infty(0)}(z) = \bar{C}
h^{(2)}_\ell(z) = \bar{C}[j_\ell(z)-i n_\ell(z)]$.
The integration of Eq.~(\ref{3.17}) for
$\bar{Y}^{\infty(1)}(z)$ proceeds as previously described.
We obtain
\begin{eqnarray}
Y^{\infty(1)}(z)/C &=& c^{(1)} h_\ell^{(1)} +
\Bigl(d^{(1)} - \mbox{Si} 2z + i\mbox{Ci} 2z
\Bigr) h_\ell^{(2)} \nonumber \\
& & \mbox{} + z^2(n_\ell j_0 - j_\ell n_0) h_0^{(1)}
\nonumber \\ & & \mbox{} +
\sum_{p=1}^{\ell-2} \biggl(\frac{1}{p} + \frac{1}{p+1}\biggr)
z^2 (n_\ell j_p - j_\ell n_p) h_p^{(1)} \nonumber \\
& & \mbox{} - \biggl[ \frac{(\ell-2)(\ell+2)}{2\ell(2\ell+1)}
+ \frac{2\ell-1}{\ell(\ell-1)} \biggr] h_{\ell-1}^{(1)}
\nonumber \\ & & \mbox{} +
\frac{(\ell-1)(\ell+3)}{2(\ell+1)(2\ell+1)} h_{\ell+1}^{(1)},
\label{3.29}
\end{eqnarray}
where $c^{(1)}$ and $d^{(1)}$ are constants of integration.

Matching to Eq.~(\ref{3.27}) involves the evaluation of
Eq.~(\ref{3.29}) in the limit $z \to \infty$, which requires
manipulations similar to the ones leading to Eq.~(\ref{3.23}).
We eventually obtain that $c^{(1)}$ is arbitrary and can be
set to zero without loss of generality, that $d^{(1)}=\pi/2$,
and that
\begin{equation}
C = (i)^{\ell+1} \bigl[ 1 + i\varepsilon (\beta_\ell + \gamma
- \ln\varepsilon) + O(\varepsilon^2) \bigr],
\label{3.30}
\end{equation}
where $\beta_\ell$ is given in Eq.~(\ref{3.24}). This completes
the determination of $Y^\infty(z)$ to first order in $\varepsilon$.

\section{Radiative multipole moments in the
         slow-motion approximation}

In this section we apply the results of
Secs.~II and III to the limiting case of a slowly
moving source. We first derive leading-order
(Newtonian) expressions for the radiative multipole
moments --- $Z^H_{\omega\ell m}$ in the frequency
domain, $\tilde{Z}^H_{\ell m}(u)$ in the time domain;
see Sec.~II B. We then improve on these expressions
by considering the corrections due to wave-propagation
(tail) effects, which are of post$^{3/2}$-Newtonian order.
These corrections do not depend on the specific details
of the source, but are a manifestation of the curvature
of spacetime outside the source. The larger, source-specific,
post-Newtonian corrections will not be considered in this
paper.

For the purpose of the following calculation we do not assume a
specific form for the stress-energy tensor $T^{\alpha\beta}$
that perturbs the black hole's gravitational field.
In particular, we do
not assume that the perturbations are produced by a moving
particle and that $T^{\alpha \beta}$ is of the form
(\ref{2.4}). [The following results
will nevertheless contain factors of $\mu^{-1}$,
the inverse of the particle's mass. This is
because such a factor has been inserted for convenience
into the definition of $Z^H_{\omega\ell m}$; see
Eq.~(\ref{2.10}). Such meaningless occurrences of
$\mu^{-1}$ can be avoided by setting $\mu \equiv 1$ throughout
this section.] We do assume, however, that the source is
slowly moving, in a sense made precise below.

\subsection{Leading-order calculation}

Let the source be characterized by some internal velocity $v$.
We demand $v \ll 1$. We assume that the source is
gravitationally bounded to
the black hole, so that $M/r \approx v^2$, where $r$ is a typical
value of the radial coordinate inside the source. Finally,
if $\omega$ is a typical frequency of the gravitational waves,
then slow motion implies $\omega r \approx v$, and
$M\omega \approx v^3$.
In terms of the variables introduced in Sec.~III, we have
$z\approx v$, $\varepsilon/z \approx v^2$, and
$\varepsilon \approx v^3$.
This implies that such results as Eqs.~(\ref{3.14}) and
(\ref{3.25}) can be used.

In the slow-motion approximation, the stress-energy tensor is
dominated by the component $\mbox{}_0 T = T_{\alpha\beta}
n^\alpha n^\beta = \frac{1}{4} \rho + O(\rho v)$, where $\rho$ is
the source's mass density. By comparison, $\mbox{}_{-1} T
= O(\rho v)$ and $\mbox{}_{-2} T = O(\rho v^2)$; these components
will be neglected. Equation (\ref{2.5}) then implies
$\mbox{}_0 T_{\omega\ell m}(r) = \frac{1}{4} \int d\Omega
\tilde{\rho}(\omega,\bbox{x}) \bar{Y}_{\ell m}(\theta,\phi)
+ O(\rho v)$, where
\begin{equation}
\tilde{\rho}(\omega,\bbox{x}) = \frac{1}{2\pi} \int
dt \rho(t,\bbox{x}) e^{i \omega t}
\label{4.1}
\end{equation}
is the mass density in the frequency domain.

The slow-motion approximation also implies that
Eqs.~(\ref{2.11}) and (\ref{2.13}) reduce to
$\kappa^H_{\omega \ell} = \frac{1}{4} (\ell-1)
\ell (\ell+1) (\ell+2) + O(v^3)$ and
$\mbox{}_0 \Gamma_{\omega\ell} = 2rd/dr + \ell(\ell+1) +
O(v)$. Equations (\ref{3.4}), (\ref{3.14}), and
(\ref{3.25}) can then be combined to give
\begin{equation}
\frac{\mbox{}_0 \Gamma_{\omega\ell} X^H_{\omega\ell}(r)}{
A^{\rm in}_{\omega\ell}} =
\frac{2(\ell+1)(\ell+2)}{(2\ell+1)!!} (-i\omega r)^{\ell+1},
\label{4.2}
\end{equation}
to leading order in $v$.
Substituting these results into Eq.~(\ref{2.10}) yields
\begin{eqnarray}
Z^H_{\omega\ell m} &=& \frac{4\pi}{(2\ell+1)!!\mu}
\biggl[ \frac{(\ell+1)(\ell+2)}{(\ell-1)\ell} \biggr]^{1/2}
(-i \omega)^\ell \nonumber \\ & & \mbox{} \times
\int d^3x\, \tilde{\rho}(\omega,\bbox{x}) r^\ell
\bar{Y}_{\ell m}(\theta,\phi),
\label{4.3}
\end{eqnarray}
the radiative moments in the frequency domain.

Finally, taking the Fourier transform of Eq.~(\ref{4.3})
we obtain, to leading order in $v$,
\begin{eqnarray}
\tilde{Z}^H_{\ell m}(u) &=& \frac{4\pi}{(2\ell+1)!!\mu}
\biggl[ \frac{(\ell+1)(\ell+2)}{(\ell-1)\ell} \biggr]^{1/2}
\biggl( \frac{d}{du} \biggr)^\ell
\nonumber \\ & & \mbox{} \times
\int d^3x\, \rho(u,\bbox{x}) r^\ell
\bar{Y}_{\ell m}(\theta,\phi),
\label{4.4}
\end{eqnarray}
the radiative moments in the (retarded) time domain. As
expected, the radiative moments of order $\ell$ are given,
up to a numerical factor,
by the $\ell$-th time derivative of the source moments,
$\int d^3 x\, \rho r^\ell \bar{Y}_{\ell m}$. This
leading-order calculation agrees with that of Sec.~V C of
Ref.~\cite{Thorne1980}. (It should be noted that we only
have obtained the mass multipole moments. To
calculate the current moments would involve keeping
terms of order $\rho v$ in our approximations.)

\subsection{Tail correction}

We now improve upon the leading-order calculation by
considering the corrections to Eq.~(\ref{4.3}) which are
due to wave-propagation effects (tails; for the
purpose of this discussion it is simpler to work in
the frequency domain). More precisely, we consider the
corrections to Eq.~(\ref{4.3}) which are of order
$M\omega \approx v^3$, the smallness parameter appearing
in the Regge-Wheeler equation. We shall ignore, for reasons
given above, the corrections of order $v$ and $v^2$.

The steps are simple. We repeat the calculation described
in the previous subsection, but with the following changes.
(i) We rewrite Eq.~(\ref{2.11}) as $\kappa^H_{\omega\ell}
= \frac{1}{4} (\ell-1)\ell(\ell+1)(\ell+2) e^{-12 i M\omega /
(\ell-1)\ell(\ell+1)(\ell+2)}$, to a fractional accuracy of
order $(M\omega)^2$. (ii) We similarly modify
$X^H_{\omega\ell}(r)$ by multiplying the leading-order
expression by $e^{2i\alpha_\ell M\omega}$, with
$\alpha_\ell$ given in Eq.~(\ref{3.15}). This
step is dictated by Eq.~(\ref{3.14});
according to our rules we may ignore the other correction
terms. (iii) We modify
$A^{\rm in}_{\omega\ell}$ by multiplying the leading-order
expression by $(1 - \pi M |\omega|) e^{-2iM\omega
(\ln 4M|\omega| - \alpha_\ell - \beta_\ell)}$, as dictated
by Eq.~(\ref{3.25}); $\beta_\ell$ is defined in Eq.~(\ref{3.24}).

After simple manipulations, the final result is that the
tail corrections to the radiative multipole moments take
the form
\begin{equation}
Z^H_{\omega\ell m} \to
Z^H_{\omega\ell m} (1 + \pi M |\omega|)
e^{2 i M\omega (\ln 4M|\omega| - \mu_\ell)}.
\label{4.5}
\end{equation}
Here, $Z^H_{\omega\ell m}$ stands for the leading-order
expression for the radiative moments. We have introduced
$\mu_\ell = \beta_\ell -
6[(\ell-1)\ell(\ell+1)(\ell+2)]^{-1}$, or
\begin{equation}
\mu_\ell = \psi(\ell-1) +
\frac{\ell^3+7\ell^2+12\ell+8}{2\ell(\ell+1)(\ell+2)},
\label{4.6}
\end{equation}
after using Eq.~(\ref{3.24}).

A result identical to Eq.~(\ref{4.5}) was first derived
for the frequency-domain quadrupole ($\ell=2$) moments by
Blanchet and Sh\"afer \cite{BlanchetShafer}, and then
generalized to moments of arbitrary order by
Blanchet \cite{BlanchetPC}. Those results were derived
within the context of post-Newtonian theory (see Secs.~I
B and G). Our results, which confirm
that of Blanchet and Sh\"afer, were obtained using a
completely different method of analysis, that of
black-hole perturbation theory. It is, of course, pleasing
that such different methods yield identical results.

Blanchet's tail correction \cite{BlanchetPC}
is identical to Eq.~(\ref{4.5})
except for a different representation of the multipole
moments (he uses a symmetric-tracefree representation),
and a different phase. Blanchet's expression substitutes
a constant $\kappa_\ell-\gamma$ in place of $\mu_\ell$,
where \cite{Blanchet}
\begin{equation}
\kappa_\ell = \sum_{k=1}^{\ell-2} \frac{1}{k} +
\frac{2\ell^2 + 5\ell + 4}{\ell(\ell+1)(\ell+2)}.
\label{4.7}
\end{equation}
It is easy to see that the two constants are related by
$\mu_\ell = \kappa_\ell - \gamma + 1/2$. Our result
therefore differs from Blanchet's by a phase factor
$e^{-i M \omega}$ independent of $\ell$. This overall
phase has no physical significance: it can
be absorbed into a shift in the origin of the
retarded time $u$.

\section{Circular orbits in the slow-motion approximation}

In this section we use the perturbation formalism
to calculate the energy radiated by a particle
in circular motion around a Schwarzschild black hole.
We perform this calculation in the slow-motion
approximation, to leading order in the velocity $v$.
We consider both the energy radiated to infinity, and
that absorbed by the black hole. Part of the calculations
presented in this section were also carried out in
paper I \cite{paperI}.
We repeat these calculations here for completeness. The
remaining calculations are presented here for the
first time. Equations (\ref{5.17}) and (\ref{5.18}) below
were quoted without derivation in a footnote
of paper III \cite{paperIII}.

\subsection{$Z^{H,\infty}_{\omega\ell m}$ for $\ell+m$ even}

We begin with the calculation of the amplitudes
$Z^{H}_{\omega\ell m}$ and $Z^\infty_{\omega\ell m}$,
to leading order in $v$ (to be defined precisely below),
for the case where $\ell+m$ is an even integer. The
case $\ell+m$ odd will be treated in subsection B. That
these cases must be considered separately will become
apparent below.

As was discussed in Sec.~IV A, the dominant component of
the stress-energy tensor in a slow-motion
approximation is $\mbox{}_0 T = T_{\alpha\beta}
n^\alpha n^\beta$. Equation (\ref{2.4}) then yields
\begin{equation}
\mbox{}_0 T = \frac{\mu}{{r_0}^2} \frac{
(u^\alpha n_\alpha)^2}{u^t}
\delta(r-r_0) \delta(\cos\theta) \delta(\phi-\Omega t).
\label{5.1}
\end{equation}
Here, $r_0$ is the radius of the circular
orbit which, without loss of generality, has been put
in the equatorial plane $\theta=\pi/2$; $u^\alpha$ is
the four-velocity, and $\Omega \equiv d\phi/dt =
(M/{r_0}^3)^{1/2}$ is the angular velocity.

We define $v$ as
\begin{equation}
v = \Omega r_0 = (M/r_0)^{1/2} = (M\Omega)^{1/3},
\label{5.2}
\end{equation}
and demand $v\ll 1$.
Up to corrections of order $M/r_0 = v^2$, Eq.~(\ref{5.1})
reduces to $\mbox{}_0 T = (\mu/4{r_0}^2) \delta(r-r_0)
\delta(\cos\theta) \delta(\phi-\Omega t)$. Substitution
into Eq.~(\ref{2.5}) and integration yields
\begin{equation}
\mbox{}_0 T_{\omega\ell m}(r) = \frac{\mu}{4{r_0}^2}\,
\mbox{}_0 Y_{\ell m}({\textstyle \frac{\pi}{2}},0)
\delta(r-r_0) \delta(\omega-m\Omega).
\label{5.3}
\end{equation}
That $\mbox{}_s T_{\omega\ell m}$, and hence
$Z^{H,\infty}_{\omega\ell m}$, is proportional to
$\delta(\omega-m\Omega)$ has been anticipated in
Eq.~(\ref{2.20}). (This result is exact, and not
a consequence of the slow-motion approximation.
See Ref.~\cite{paperI} for details.)  We shall now
factor out this $\delta$ function, and work with the
quantities $Z^{H,\infty}_{\ell m}$ defined in
Eq.~(\ref{2.20}).

Substituting Eqs.~(\ref{4.2}) and (\ref{5.3}) into
(\ref{2.10}) and integrating, we obtain
\begin{eqnarray}
Z^{H}_{\ell m} &=& 4\pi (-i)^\ell
\frac{m^\ell}{(2\ell+1)!!}
\biggl[ \frac{(\ell+1)(\ell+2)}{(\ell-1)\ell} \biggr]^{1/2}
\nonumber \\ & & \mbox{} \times\,
\mbox{}_0 Y_{\ell m}({\textstyle \frac{\pi}{2}},0)
v^\ell.
\label{5.4}
\end{eqnarray}
Equation (\ref{5.4}) is valid up to fractional corrections
of order $v^2$ (see Refs.~\cite{TagoshiSasaki,paperI} for
details.)

A similar calculation yields $Z^\infty_{\ell m}$. We
first use the results of Sec.~III E to obtain
$X^\infty_{\omega \ell}(r) = (i)^{\ell+1} \omega r
h_\ell^{(1)}(\omega r)$, where $h_\ell^{(1)}$ is the
spherical Hankel function, to leading order in $M\omega
= O(v^3)$. Then, up to fractional corrections of order
$(\omega r)^2 = O(v^2)$, we have
$X^\infty_{\omega\ell}(r) =
(i)^\ell (2\ell-1)!! (\omega r)^{-\ell}$. Next, we
use the leading-order expressions for
$\mbox{}_0 \Gamma_{\omega \ell}$ and
$A^{\rm in}_{\omega \ell}$, Eqs.~(\ref{2.13}) and
(\ref{3.25}), to obtain
\begin{equation}
\frac{\mbox{}_0 \Gamma_{\omega\ell} X^\infty_{\omega \ell(r)}}{
A^{\rm in}_{\omega \ell}} = - 2^{\ell+2} i
\frac{(\ell-1)\ell(\ell-2)!(\ell+2)!}{(2\ell)!(2\ell+1)}
\frac{(M\omega)^{\ell+1}}{(\omega r)^\ell},
\label{5.5}
\end{equation}
to leading order in $v$. Finally, substitution of
Eqs.~(\ref{5.3}) and (\ref{5.5}) into (\ref{2.10})
yields
\begin{eqnarray}
Z^\infty_{\ell m} &=& -2^{\ell-3} \pi
\frac{\bigl[(\ell-1)\ell\bigr]^{3/2}
\bigl[(\ell+1)(\ell+2)\bigr]^{1/2}}
{m^3 (2\ell)! (2\ell+1)}
\nonumber \\ & & \mbox{} \times
(\ell-2)! (\ell+2)! \,
\mbox{}_0 Y_{\ell m}({\textstyle \frac{\pi}{2}},0)
v^{2\ell - 7},
\label{5.6}
\end{eqnarray}
to leading order in $v$.

An explicit expression for
$\mbox{}_0 Y_{\ell m}({\textstyle \frac{\pi}{2}},0)$
can be obtained by expressing the spherical harmonics
in terms of the associated Legendre polynomials, and
using the fact that $P_\ell^m (0) =
(-1)^{\frac{1}{2} (\ell-m)}
(\ell+m-1)!!/(\ell-m)!!$ if $\ell+m$ is even, and
$P_\ell^m (0) = 0$ if $\ell+m$ is odd \cite{Arfken}.
Thus,
\begin{eqnarray}
\mbox{}_0 Y_{\ell m}({\textstyle \frac{\pi}{2}},0) &=&
(-1)^{\frac{1}{2}(\ell+m)}
\biggl[ \frac{2\ell+1}{4\pi} \biggr]^{1/2}
\nonumber \\ & & \mbox{} \times
\frac{ \bigl[ (\ell-m)!(\ell+m)! \bigr]^{1/2}}{(\ell-m)!!
(\ell+m)!!}
\label{5.7}
\end{eqnarray}
if $\ell + m$ is even, and
$\mbox{}_0 Y_{\ell m}({\textstyle \frac{\pi}{2}},0) = 0$
otherwise. We therefore see that the dominant, $s=0$,
contribution to $Z^{H,\infty}_{\ell m}$ vanishes
identically if $\ell+m$ is odd.

\subsection{$Z^{H,\infty}_{\omega\ell m}$ for $\ell+m$ odd}

If $\ell+m$ is odd the dominant contribution to
$Z^{H,\infty}_{\ell m}$ comes from the component
$\mbox{}_{-1}T = T_{\alpha\beta} n^\alpha \bar{m}^\beta$ of
the stress-energy tensor. A calculation analogous
to that of the previous subsection yields $\mbox{}_{-1} T =
(i\mu v/ 2\sqrt{2} {r_0}^2) \delta(r-r_0) \delta(\cos\theta)
\delta(\phi-\Omega t)$, so that
\begin{equation}
\mbox{}_{-1} T_{\omega\ell m}(r) =
\frac{i \mu v}{2\sqrt{2} {r_0}^2}\,
\mbox{}_{-1} Y_{\ell m}({\textstyle \frac{\pi}{2}},0)
\delta(r-r_0) \delta(\omega - m \Omega).
\label{5.8}
\end{equation}
{}From Eq.~(\ref{2.13}) we have
$\mbox{}_{-1} \Gamma_{\omega \ell} = -\ell(\ell+1)(rd/dr+1)$.
It follows that
\begin{equation}
\frac{
\mbox{}_{-1} \Gamma_{\omega \ell} X^H_{\omega\ell}(r)}{
A^{\rm in}_{\omega \ell}} = -2 (-i)^{\ell+1}
\frac{\ell(\ell+1)(\ell+2)}{(2\ell+1)!!} (\omega r)^{\ell+1}
\label{5.9}
\end{equation}
and
\begin{eqnarray}
\frac{
\mbox{}_{-1} \Gamma_{\omega \ell} X^\infty_{\omega\ell}(r)}{
A^{\rm in}_{\omega \ell}} &=& -2^{\ell+2} i
\frac{(\ell-1)\ell(\ell+1)}{(2\ell)!
(2\ell+1)} (\ell-2)!(\ell+2)!
\nonumber \\ & & \mbox{} \times
(M\omega)^{\ell+1} (\omega r)^{-\ell}.
\label{5.10}
\end{eqnarray}
Here, $X^H_{\omega\ell}(r)$,
$X^\infty_{\omega\ell}(r)$, and $A^{\rm in}_{\omega\ell}$
were approximated as in the previous subsection.

We arrive at
\begin{eqnarray}
Z^H_{\ell m} &=& 8\pi(-i)^{\ell+1}
\biggl[ \frac{\ell+2}{\ell-1} \biggr]^{1/2}
\frac{m^\ell}{(2\ell+1)!!}
\nonumber \\ & & \mbox{} \times
\mbox{}_{-1} Y_{\ell m}({\textstyle \frac{\pi}{2}},0)
v^{\ell+1}
\label{5.11}
\end{eqnarray}
and
\begin{eqnarray}
Z^\infty_{\ell m} &=& -2^{\ell-2} i \pi
\frac{\bigl[(\ell-1)(\ell+2)\bigr]^{1/2}
}{m^3 (2\ell)! (2\ell+1)} (\ell+1) \ell! (\ell+2)!
\nonumber \\ & & \mbox{} \times
\mbox{}_{-1} Y_{\ell m}({\textstyle \frac{\pi}{2}},0)
v^{2\ell-6},
\label{5.12}
\end{eqnarray}
to leading order in $v$.

An explicit expression for
$\mbox{}_{-1} Y_{\ell m}({\textstyle \frac{\pi}{2}},0)$
can be obtained as follows. First, we calculate the
$s=-1$ spherical harmonics \cite{Goldbergetal} using
$\mbox{}_{-1} Y_{\ell m}(\theta,\phi) =
-[\ell(\ell+1)]^{-1/2} (\partial_\theta - i \csc\theta
\partial_\phi) \mbox{}_0 Y_{\ell m}(\theta,\phi)$. We
then express the $s=0$ harmonics in terms of the
associated Legendre polynomials. Since $P_\ell^m(0)=0$
when $\ell+m$ is odd, we obtain
$\mbox{}_{-1} Y_{\ell m}({\textstyle \frac{\pi}{2}},0) =
(-1)^m [(2\ell+1)/4\pi\ell(\ell+1)]^{1/2} [(\ell-m)!/
(\ell+m)!]^{1/2} P_\ell^{m \prime}(0)$ if $\ell+m$ is
odd. (Here, a prime denotes differentiation with
respect to the argument.) Finally,  use of the
relation \cite{Arfken} $P_\ell^{m\prime}(0) =
\frac{1}{2} P_{\ell}^{m+1}(0) - \frac{1}{2}
(\ell+m)(\ell-m+1) P_\ell^{m-1}(0)$ yields
\begin{eqnarray}
\mbox{}_{-1} Y_{\ell m}({\textstyle \frac{\pi}{2}},0) &=&
(-1)^{\frac{1}{2} (\ell+m-1)}
\biggl[ \frac{2\ell+1}{4\pi\ell(\ell+1)} \biggr]^{1/2}
\nonumber \\ & & \mbox{} \times
\frac{(\ell-m)!!(\ell+m)!!}{
\bigl[(\ell-m)!(\ell+m)!\bigr]^{1/2}}
\label{5.13}
\end{eqnarray}
if $\ell+m$ is odd. For $\ell+m$ even
$\mbox{}_{-1} Y_{\ell m}({\textstyle \frac{\pi}{2}},0)$
is nonzero, but its expression will not be needed.

\subsection{Energy radiated}

It is now a straightforward task to calculate the
rate at which the gravitational waves remove energy
from the system.
We use the results of the preceding subsections,
together with Eqs.~(\ref{2.21})--(\ref{2.23}).
For convenience we shall normalize our expressions for
$\dot{E}^\infty$ and $\dot{E}^H$ to the
quadrupole-formula expression, $\dot{E}_{\rm QF} =
(32/5) (\mu/M)^2 v^{10}$ \cite{Thorne1987}.
We therefore define the
numbers $\eta^{H,\infty}_{\ell m}$ such that
\begin{equation}
\dot{E}^{H,\infty} = \frac{32}{5}
\Bigl( \frac{\mu}{M} \Bigr)^2
v^{10} \sum_{\ell m}
{\textstyle \frac{1}{2}}
\eta^{H,\infty}_{\ell m}.
\label{5.14}
\end{equation}
The factor of $1/2$ is inserted for convenience
because of the symmetry $\eta^{H,\infty}_{\ell,-m} =
\eta^{H,\infty}_{\ell m}$ which can be derived from
$\bar{Z}^{H,\infty}_{-\omega,\ell,-m} =
(-1)^\ell Z^{H,\infty}_{\omega\ell m}$
(see paper III \cite{paperIII} for a derivation).

Calculation yields, to leading order in $v$,
\begin{equation}
\eta^\infty_{\ell m} = \frac{5\pi}{4}
\frac{m^{2(\ell+1)}}{(2\ell+1)!!^2}
\frac{(\ell+1)(\ell+2)}{(\ell-1)\ell}
\bigl[\mbox{}_{0} Y_{\ell m}
({\textstyle \frac{\pi}{2}},0)\bigr]^2
v^{2(\ell-2)}
\label{5.15}
\end{equation}
if $\ell+m$ is even, and
\begin{equation}
\eta^\infty_{\ell m} = 5\pi
\frac{m^{2(\ell+1)}}{(2\ell+1)!!^2}
\frac{(\ell+2)}{(\ell-1)}
\bigl[\mbox{}_{-1} Y_{\ell m}
({\textstyle \frac{\pi}{2}},0)\bigr]^2
v^{2(\ell-1)}
\label{5.16}
\end{equation}
if $\ell+m$ is odd.

Calculation also yields, to leading order in $v$,
\begin{equation}
\eta^H_{\ell m} = \biggl[
\frac{2 \ell!}{m^\ell} \biggr]^2 v^{2(\ell+2)}\,
\eta^\infty_{\ell m}
\label{5.17}
\end{equation}
if $\ell + m$ is even, and
\begin{equation}
\eta^H_{\ell m} = \biggl[
\frac{2 (\ell+1)!}{\ell m^\ell} \biggr]^2
v^{2(\ell+2)}\, \eta^\infty_{\ell m}
\label{5.18}
\end{equation}
if $\ell+m$ is odd.

\subsection{Black-hole absorption}

In a slow-motion approximation, the dominant
contribution to the energy radiated comes from
the $\ell=2$, $|m|=2$ (mass quadrupole)
terms in Eq.~(\ref{5.14}). The dominant
contribution to the black-hole absorption
is given by $\eta^H_{22} = v^8 \eta^\infty_{22}
= v^8$. Therefore
\begin{equation}
\dot{E}^H/\dot{E}^\infty = v^8
\bigl[1 + O(v^2)\bigr].
\label{5.19}
\end{equation}
(That the correction must be of order $v^2$ can be
established by direct calculation.)

In a slow-motion approximation, the total
energy radiated, $\dot{E} = \dot{E}^\infty
+ \dot{E}^H$, can be expressed as a
post-Newtonian expansion of the form
\begin{eqnarray}
\dot{E} &=& \dot{E}_{\rm QF}
\Bigl[ 1 + O(v^2) + O(v^3) + O(v^4) + O(v^5)
\nonumber \\ & & \mbox{}
+ O(v^6) + O(v^6 \ln v) + O(v^7) + O(v^8)
\nonumber \\ & & \mbox{}
+ O(v^8 \ln v) + \cdots \Bigl].
\label{5.20}
\end{eqnarray}
The first three terms of this expansion,
through $O(v^3)$, were calculated in paper I
\cite{paperI}; all other terms were calculated by
Tagoshi and Sasaki \cite{TagoshiSasaki}. (See
Sec.~I B and D for a more detailed discussion.)

We therefore see that the contribution $\dot{E}^H$
to the total energy radiated occurs at quite a high
order in the post-Newtonian expansion: $O(v^8)$,
or post$^4$-Newtonian order. What is more, the
black-hole absorption is also a small contribution to
the term of order $v^8$. The coefficient of the $O(v^8)$
term in the post-Newtonian expansion of $\dot{E}^\infty$
was calculated, in the limit of small mass ratios, by
Tagoshi and Sasaki \cite{TagoshiSasaki}.
They find that it is approximately
equal to $-117.5044$. The coefficient of the $O(v^8)$ term
in the expansion of $\dot{E}^H$ is given in
Eq.~(\ref{5.19}): it is equal to unity. The black-hole
absorption therefore contributes less than one
percent of the term of order $v^8$ in Eq.~(\ref{5.20}).

The black-hole absorption is a small effect indeed.

\section*{Acknowledgments}

Conversations with Luc Blanchet and Clifford Will were
greatly appreciated. Eric Poisson's work was supported
in part by the National Science Foundation under Grant
No.~PHY 92-22902 and the National Aeronautics and Space
Administration under Grant No.~NAGW 3874. Misao
Sasaki's work was supported in part by the Japanese
Grant-in-Aid for Scientific Research on Priority Areas
of the Ministry of Education, Science, and Culture,
No.~04234104.
\appendix
\section{Derivation of Eq.~(2.10)}

The source term to the right-hand side of
the inhomogeneous Teukolsky equation,
Eq.~(\ref{2.2}), is given explicitly by
\cite{paperI,SasakiNakamura}
\begin{equation}
T_{\omega \ell m}(r) = 2\pi \sum_s \mbox{}_s p_\ell
\, \mbox{}_s D_{\omega}
\, \mbox{}_s T_{\omega \ell m} (r),
\label{A.1}
\end{equation}
where the constants $\mbox{}_s p_\ell$ are listed in
Eq.~(\ref{2.12}), the functions $\mbox{}_s T_{\omega
\ell m}(r)$ given in Eq.~(\ref{2.5}), and where
\begin{equation}
\mbox{}_s D_\omega = \left\{
\begin{array}{ll}
r^4 & s=0 \\
r^2 f {\cal L} r^3 f^{-1} & s=-1 \\
r f {\cal L} r^4 f^{-1} {\cal L} r & s=-2
\end{array}
\right.
\label{A.2}
\end{equation}
are differential operators. Here, ${\cal L}
= f d/dr + i\omega = d/dr^* + i \omega$.

Equation (\ref{2.2}) is integrated by constructing
a Green's function from two linearly independent solutions
to the homogeneous equation \cite{Detweiler}.
These are denoted $R^H_{\omega
\ell}(r)$ and $R^\infty_{\omega\ell}(r)$, and have the following
asymptotic behaviors: $R^H_{\omega\ell}(r\to 2M) \sim
(\omega r)^4 f^2 e^{-i \omega r^*}$; $R^H_{\omega\ell}
(r\to\infty) \sim Q^{\rm in}_{\omega\ell} (\omega r)^{-1}
e^{-i \omega r^*} + Q^{\rm out}_{\omega\ell} (\omega r)^{3}
e^{i \omega r^*}$ ($Q^{\rm in}_{\omega\ell}$ and
$Q^{\rm out}_{\omega\ell}$ are constants);
$R^\infty_{\omega\ell} (r\to\infty) \sim (\omega r)^3
e^{i\omega r^*}$. A straightforward application of the
general theory of Green's functions then shows that the
solution to the inhomogeneous Teukolsky equation reduces
to Eqs.~(\ref{2.8}) and (\ref{2.9}), with
\begin{eqnarray}
Z^{H,\infty}_{\omega \ell m} &=&
\Bigl[2 i \mu \omega^2 Q^{\rm in}_{\omega \ell}
\Bigr]^{-1} \nonumber \\
& & \mbox{} \times
\int_{2M}^\infty dr\, r^{-4} f^{-2}
R^{H,\infty}_{\omega \ell}(r) T_{\omega\ell m}(r).
\label{A.3}
\end{eqnarray}

Our goal in this Appendix is to rewrite Eq.~(\ref{A.3}) into the
form (\ref{2.10}). The first step is to substitute
Eq.~(\ref{A.1}) into (\ref{A.3}) and to introduce the adjoint
operators $\mbox{}_s D_\omega^\dagger$ such that this
can be written in the equivalent form
\begin{eqnarray}
Z^{H,\infty}_{\omega \ell m} &=&
\pi \Bigl[i \mu \omega^2 Q^{\rm in}_{\omega \ell} \Bigr]^{-1}
\sum_s \mbox{}_s p_\ell \nonumber \\
& & \mbox{} \times
\int_{2M}^\infty dr\, r^{-4} f^{-2}
\mbox{}_s T_{\omega\ell m}(r)\,
\mbox{}_s D_\omega^\dagger
R^{H,\infty}_{\omega \ell}(r).
\label{A.4}
\end{eqnarray}
Equation (\ref{A.4}) can be obtained from (\ref{A.3}) by
performing a number of integration by parts, and the
adjoint operators as thus determined. They are found
to be given by
\begin{equation}
\mbox{}_s D_\omega^\dagger = \left\{
\begin{array}{ll}
r^4 & s=0 \\
-r^7 \bar{{\cal L}} r^{-2} & s=-1 \\
r^5 f \bar{{\cal L}} r^4 f^{-1}
\bar{{\cal L}} r^{-3} & s=-2
\end{array}
\right. ,
\label{A.5}
\end{equation}
where $\bar{\cal L} = d/dr^*-i\omega$. It is useful to note
that the operator adjoint to $\cal L$ is ${\cal L}^\dagger
= -r^4 f \bar{\cal L} (r^4 f)^{-1}$. With this known,
Eq.~(\ref{A.5}) can be obtained directly from (\ref{A.2}).

The next step is to relate the functions
$R^{H,\infty}_{\omega \ell}(r)$ to the
Regge-Wheeler functions $X^{H,\infty}_{\omega
\ell}(r)$, introduced
in Eq.~(\ref{2.7}). A straightforward calculation
(see Ref.~\cite{GRR} for details) shows that these
are related by the Chandrasekhar transformation
\cite{ChandraTransf}
\begin{equation}
R^{H,\infty}_{\omega\ell}(r) =
\chi^{H,\infty}_{\omega \ell} C_\omega
X^{H,\infty}_{\omega\ell}(r),
\label{A.6}
\end{equation}
where
\begin{eqnarray}
\chi^H_{\omega\ell} &=& \frac{16(1-2iM\omega)(1-4iM\omega)}{
(\ell-1)\ell(\ell+1)(\ell+2) - 12iM\omega}\, (M\omega)^3,
\nonumber \\
& & \label{A.7} \\
\chi^\infty_{\omega\ell} &=& - \frac{1}{4}, \nonumber
\end{eqnarray}
and where
\begin{equation}
C_{\omega} = \omega r^2 f {\cal L} f^{-1} {\cal L} r
\label{A.8}
\end{equation}
is a second-order differential operator. Equation (\ref{A.6})
implies
\begin{equation}
Q^{\rm in}_{\omega \ell} = -4 (1-2iM\omega)(1-4iM\omega)
(M\omega)^3 A^{\rm in}_{\omega \ell};
\label{A.9}
\end{equation}
the constant $A^{\rm in}_{\omega\ell}$ was introduced in
Eq.~(\ref{2.7}).

In order to re-express $Z^{H,\infty}_{\omega\ell m}$ as
in Eq.~(\ref{2.10}) we must operate with $\mbox{}_s
D_\omega^\dagger C_\omega$ on the Regge-Wheeler functions.
These operations involve successive applications of
$\cal L$, and the Regge-Wheeler equation (\ref{2.6}) is
substituted as often as necessary to simplify the
expressions. To carry out these manipulations efficiently, it is
useful to first develop the algebra of the $\cal L$ and
$\bar{\cal L}$ operators. We now list the most useful rules.

When acting on any solution
$X_{\omega \ell}(r)$ of the Regge-Wheeler equation, two
successive applications of $\cal L$ or $\bar{\cal L}$
reduce to:
\begin{eqnarray}
{\cal L} {\cal L} &=& 2i\omega {\cal L} + V(r),
\nonumber \\
\bar{\cal L} {\cal L} &=& {\cal L} \bar{\cal L} = V(r),
\label{A.10} \\
\bar{\cal L} \bar{\cal L} &=& -2i\omega \bar{\cal L} + V(r).
\nonumber
\end{eqnarray}
Here, $V(r)$ is the Regge-Wheeler potential, introduced in
Eq.~(\ref{2.6}). When carrying out the calculations it is also
useful to invoke the commutation relations $[{\cal L},g(r)]
= [\bar{\cal L},g(r)] = f dg/dr$, where $g$ is any function
of $r$.

Using these rules the calculations are straightforward.
The final answer takes the form (\ref{2.10}) if we
define $\mbox{}_s \Gamma_{\omega \ell} =
\omega^{-1} r^{-5} \mbox{}_s D_\omega^\dagger C_\omega$,
and $\kappa^{H,\infty}_{\omega\ell} = -
Q^{\rm in}_{\omega\ell}/\chi^{H,\infty}_{\omega\ell}
A^{\rm in}_{\omega\ell}$.


\begin{references}
\bibitem{foot} For $\omega < 0$ one defines $z=|\omega|r$,
        $\varepsilon = 2M|\omega|$, $z^* = z + \varepsilon
        \ln(z-\varepsilon)$, $Y^H(z) = z^{-1} e^{-i
        \varepsilon \ln(z-\varepsilon)} X^H(z)$, and
        $Y^\infty(z) = z^{-1} e^{i\varepsilon \ln(z-
        \varepsilon)} X^\infty(z)$. Then
        Eqs.~(3.11)--(3.14), (3.18), (3.20), (3.22),
        (3.25), and (3.28)--(3.30) continue to hold
        provided that their right-hand sides be replaced
        by their complex conjugate.
\pagebreak
\end{references}
\end{document}